\documentclass[aps,preprintnumbers,amsmath,amssymb,prd,superscriptaddress,longbibliography,twocolumn]{revtex4-2}
\usepackage{mwe}
\usepackage{dcolumn}
\usepackage{color}
\usepackage{xcolor}
\usepackage[caption=false]{subfig}
\usepackage{feynmp}
\usepackage{feynmp-auto}
\usepackage{float}
\usepackage{systeme,mathtools}
\usepackage{easyReview}
\usepackage{amsmath,amssymb,amsfonts,empheq,physics,float,mathrsfs,wrapfig,indentfirst,enumitem}

\def\comment#1{}

\newcommand{\nc}{\newcommand}

\nc{\scs}{\scriptstyle}
\nc{\setval}{\fmfset{wiggly_len}{3mm} \fmfset{arrow_len}{1.5mm}
	\fmfset{arrow_ang}{13} \fmfset{dash_len}{1.5mm}\fmfpen{0.125mm}
	\fmfset{dot_size}{2thick}}

\usepackage{bm,latexsym,mathrsfs,enumerate,color}
\usepackage[mathcal]{euscript}
\usepackage[breaklinks=true,unicode=true,urlcolor = blue,colorlinks = true,citecolor = blue,linkcolor = blue]{hyperref}
\usepackage{graphicx}
\usepackage{todonotes}
\usepackage{wrapfig}
\usepackage{easyReview}
\usepackage{mathbbol}
\usepackage{stmaryrd}

\def\slashchar#1{\setbox0=\hbox{$#1$}           
	\dimen0=\wd0                                 
	\setbox1=\hbox{/} \dimen1=\wd1               
	\ifdim\dimen0>\dimen1                        
	\rlap{\hbox to \dimen0{\hfil/\hfil}}      
	#1                                        
	\else                                        
	\rlap{\hbox to \dimen1{\hfil$#1$\hfil}}   
	/                                         
	\fi}                                         %

\DeclareMathAlphabet\mathbfcal{OMS}{cmsy}{b}{n}
\DeclareSymbolFontAlphabet{\amsmathbb}{AMSb}%

\begin{document}
	
\title{Momentum Space Entanglement from the Wilsonian Effective Action}

\author{Matheus H. Martins Costa}
\affiliation{Institute for Theoretical Solid State Physics, IFW Dresden, Helmholtzstr. 20, 01069 Dresden, Germany}
\affiliation{Instituto de F\'{i}sica Te\'orica, Universidade Estadual Paulista,
Rua Dr. Bento Teobaldo Ferraz, 271 - Bloco II, 01140-070 S\~ ao Paulo, SP, Brazil}

\author{Jeroen van den Brink}
\affiliation{Institute for Theoretical Solid State Physics, IFW Dresden, Helmholtzstr. 20, 01069 Dresden, Germany}
\affiliation{Institute for Theoretical Physics and W\"urzburg-Dresden Cluster of Excellence ct.qmat, TU Dresden, 01069 Dresden, Germany}

\author{Flavio S. Nogueira}
\affiliation{Institute for Theoretical Solid State Physics, IFW Dresden, Helmholtzstr. 20, 01069 Dresden, Germany}

\author{Gast\~ ao I. Krein}
\affiliation{Instituto de F\'{i}sica Te\'orica, Universidade Estadual Paulista,
Rua Dr. Bento Teobaldo Ferraz, 271 - Bloco II, 01140-070 S\~ ao Paulo, SP, Brazil}

\begin{abstract}
	The entanglement between momentum modes of a quantum field theory at different scales is not as well studied as its counterpart in real space, despite the natural connection with the Wilsonian idea of integrating out the high-momentum degrees of freedom. Here, we push such connection further by developing a novel method to calculate the Rényi and entanglement entropies between slow and fast modes, which is based on the Wilsonian effective action at a given scale. This procedure is applied to the perturbative regime of some scalar theories, comparing the lowest-order results with those from the literature and interpreting them in terms of Feynman diagrams. This method is easily generalized to higher-order or nonperturbative calculations. It has the advantage of avoiding matrix diagonalizations of other techniques. 
\end{abstract}
\maketitle

\section{Introduction}

The application of information concepts to the study of quantum field theories (QFTs) is nowadays a well-established and fruitful line of research: from investigations on the connection between entanglement of regions of space and black hole entropy \cite{bomb,Srednicki} to applications in holography \cite{Ryu,nishioka}, passing through derivations of emergent symmetries in low-energy scattering \cite{Kaplan,Low:2021ufv}, understanding the entanglement structure of field theories has brought new insights on the properties of these systems. In particular, entanglement is increasingly seen as being of key relevance to quantum phase transitions \cite{Vidal}, conformal field theories (CFTs) in general \cite{cc,Sachdev}, and even as a way to characterize topological phases \cite{Kitaev:2005dm,XG}. 

Most of these studies have the common feature that they mainly focus on the properties of \textit{real-space} entanglement, i.e., on the entanglement between a region of space and its complement or between separate regions. Such a preference for entanglement in configuration space is often justified by arguing that observables typically measured in a QFT are local (effectively supported in a bounded region), and thus spatial correlations are directly accessible, having a straightforward physical interpretation. This is of course correct, but does not take into account the fact that actual measurements made in the lab have a finite resolution, so that they only detect modes up to a certain momentum scale. This is associated with the physics behind the idea of renormalization \cite{wilson,goldenfeld}. Furthermore, since renormalization (more specifically, Wilsonian renormalization) is naturally formulated in terms of momenta above and below a certain scale, there may be a lot to learn about QFTs and the previously mentioned topics by studying \textit{momentum-space} entanglement and its connection to the renormalization group (RG). After all, RG trajectories are of paramount importance to the modern understanding of the phase structure of field theories. 

It is important to note that there are studies of renormalization in the context of entanglement of spatial regions, see, for example, Refs. \cite{Miqueleto:2020qhh,Klco:2021biu}, Refs. \cite{Iso1,Iso2} which make connections with the Wilsonian effective action (still in a real-space context), and section VIII of the review article  Ref. \cite{nishioka}. There are also explorations of entanglement in momentum space such as \cite{Balasubramanian:2011wt, Agon:2014uxa,Agon:2017oia}, the first being one of the main references in this paper, Refs.  \cite{Hsu:2012gk,Flynn:2022tbj} (both for fermions at finite density, with the latter using a Gaussian approximation), \cite{Kawamoto:2021rox} (application to theories in a noncommutative space), \cite{Grignani:2016igg,Peschanski:2016hgk} for connections with particle scattering and the numerical analyses in \cite{Lundgren1,Lundgren2}. This partition was also investigated in relation to holography in \cite{Balasubramanian:2012hb} and the recent work \cite{pedraza}, where a generalization of was the so-called ``entanglement wedge" was proposed for momentum space. Nevertheless, this line of research is still in its (relative) infancy and the connection between renormalization and momentum-space entanglement is far from fully understood.

A first step towards such understanding was given in Ref. \cite{Balasubramanian:2011wt}, where it was pointed out that a reduced density matrix for low-momentum degrees of freedom at a scale $\mu$ in the vacuum of a QFT is naturally associated with the Wilsonian effective action $S_\mu[\phi_{\Vec{k}}]$ (obtained from the bare action $S[\phi_{\Vec{k}}]$ of the theory by integrating out all field modes with momentum $\Vec{k}$ such that $|\Vec{k}|\geq \mu$ \cite{wilson}), with matrix elements given by the path integral in the zero temperature limit: 
\begin{equation}
\label{density}
      \bra{\varphi_{\Vec{k}}}\rho_\mu\ket{\Tilde{\varphi}_{\Vec{k}}} = \lim_{\beta\to\infty} \frac{1}{Z(\beta)} \int_{\phi_{\Vec{k}}(0)=\varphi_{\Vec{k}}}^{\phi_{\Vec{k}}(\beta)=\Tilde{\varphi}_{\Vec{k}}}\mathcal{D}\phi_{\Vec{k}}(\tau)e^{-S^\beta_\mu}.
\end{equation}
However, the relation above was actually not used in Ref. \cite{Balasubramanian:2011wt} to obtain the entanglement entropy between low and high momentum degrees of freedom, relying instead on a Hamiltonian formalism valid only in the perturbative regime and whose connection to the Wilsonian renormalization is not obvious.

With this, our goal in this paper is to develop a new method for deriving the entanglement and Rényi entropies directly from the effective action and which has also the advantage of being well-defined nonperturbatively. 
The structure of the paper is, then, as follows. In Section \ref{lowmomentum} we review how the reduced density matrix $\rho_\mu$ is obtained from the restriction of observables to a low-momentum sector and how this automatically connects $\rho_\mu$ to $S_\mu[\phi_{\Vec{k}}]$. Then we proceed to constructing our method for calculating $\Tr\rho_\mu^n$ for $n$ integer (valid entanglement measures on their own) based on equation (\ref{density}), obtaining in this way the Rényi entropies $H_n(\rho_\mu) \equiv \frac{1}{1-n}\log\Tr\rho_\mu^n$ for any $n$, as well as the entanglement entropy $S_{EE}\equiv -\Tr\rho_\mu\log\rho_\mu$ via the replica trick \cite{cc,nishioka}. This novel technique depends on the fact that the Wilsonian integration of fast modes generate effective actions nonlocal in time (this will be made more precise later on). This turns out to be an intuitive property which usually does not need to be taken into account when calculating correlation functions, but becomes crucial when deriving the entanglement properties of the theory. Thus, part of Section \ref{lowmomentum} is also dedicated to discussing how this nonlocality is a necessary requirement to obtaining nonzero entropy.

In Section \ref{applications} we apply the method to calculate entanglement measures in cases for which analytical calculations are mostly possible, and whose details are found in Appendices \ref{a1}, \ref{a2} and \ref{a3}. As a first application of the method, we calculate the entanglement between coupled harmonic oscillators in the perturbative regime (Section \ref{harmonic}), in which case we find an agreement with \cite{nishioka}. Then, we move on to more complex examples and calculate the momentum-space entanglement for the scalar $\phi^3$ (Section \ref{phi3}) and $\phi^4$ (Section \ref{phi4}) theories up to the lowest non-trivial order in perturbation theory, and reproduce the results from \cite{Balasubramanian:2011wt}. In doing so we are also able to connect these entropies to specific Feynman diagrams, which suggests that Feynman rules for entanglement may be defined at all orders, a possibility left for further study. Section \ref{perturb} concludes the paper by explaining how the $n\to1$ limit of the replica trick must be dealt with in perturbation theory in order to get the correct results for the entanglement entropy.


\section{Density matrix and the replica trick in momentum space}
\label{lowmomentum}

The idea of restricting observables of a QFT to a ``low-momentum" sector which extends only up to a cutoff $\mu$ has a very natural realization within the path integral formalism, which we will use in this Section to define density matrices in momentum space, and from these calculate entanglement entropies.

Note, however, that the technique developed here is very general and can be applied to other contexts as long as a path integral definition of a density matrix is available, though in other cases a strict connection with the RG is not guaranteed.

\subsection{Reduced density matrix for low-momentum degrees of freedom}

The usual construction of the path integral, reviewed in Ref. \cite{nishioka}, naturally defines a way of representing the matrix elements of a density operator $\rho$, since those are transition amplitudes and thus susceptible to Feynman's technique. 

In particular, given a QFT with Euclidean action $S[\phi]$ and field operators collectively denoted by $\hat{\phi}(\Vec{x})$, whose Fourier transforms are $\hat{\phi}_{\Vec{k}}$, the matrix elements of the vacuum density operator $\rho$ in the momentum representation are given by \cite{Balasubramanian:2011wt,nishioka},
\begin{equation}
\label{groundstate}
      \bra{\varphi_{\Vec{k}}}\rho\ket{\Tilde{\varphi}_{\Vec{k}}} = \lim_{\beta\to\infty} \frac{1}{Z(\beta)} \int_{\phi_{\Vec{k}}(0)=\varphi_{\Vec{k}}}^{\phi_{\Vec{k}}(\beta)=\Tilde{\varphi}_{\Vec{k}}}\mathcal{D}\phi_{\Vec{k}}(\tau)e^{-S^\beta}.
\end{equation}
This leads to the usual expression for calculating the ground state expectation value of any observable $\mathbb{O}$, which is given in momentum space by some function $\mathbb{O} = \mathbb{O}\left(\phi_{\Vec{k}},i\frac{\delta}{\delta\phi_{\Vec{k}}}\right)$ (see Ref. \cite{jackiw}),
\begin{equation}
\label{expectation}
    \langle\mathbb{O}\rangle = \frac{1}{Z}\int\mathcal{D}\phi_{\Vec{k}} \mathbb{O}\left(\phi_{\Vec{k}},i\frac{\delta}{\delta\phi_{\Vec{k}}}\right)e^{-S[\phi]}.
\end{equation}

Now, since any measuring device that can be built in a lab is only able to resolve phenomena up to a certain momentum scale, denoted here by $\mu$, the corresponding observables are described only by functionals of $\phi_{\Vec{k}}$ such that $|\Vec{k}|\leq \mu$.

Consequently, the expectation value of such a low-momentum observable is,
\begin{equation}
\label{lowmomentumobs}
    \begin{split}
        \langle\mathbb{O}\rangle &= \frac{1}{Z}\int\mathcal{D}\phi_{\Vec{k}} \mathbb{O}(\phi_{\Vec{k}},\frac{\delta}{\delta\phi_{\Vec{k}}})e^{-S[\phi_{\Vec{k}}]} \\
        &= \frac{1}{Z}\int\prod_{|\Vec{k}|\leq\mu}\mathcal{D}\phi_{\Vec{k}} \mathbb{O}(\phi_{\Vec{k}},\frac{\delta}{\delta\phi_{\Vec{k}}})e^{-S_\mu[\phi_{|\Vec{k}|\leq\mu}]},
    \end{split}
\end{equation}
where the Wilsonian effective action at scale $\mu$, denoted by $S_\mu[\phi_{|\Vec{k}|\leq\mu}]$, is defined as usual \cite{wilson} by,
\begin{equation}
\label{effaction}
    e^{-S_\mu[\phi_{|\Vec{k}|\leq\mu}]}\equiv \int\prod_{|\Vec{k}|>\mu}\mathcal{D}\phi_{\Vec{k}}e^{-S[\phi_{\Vec{k}}]},
\end{equation}
and is \textit{automatically} obtained, since the observable has no dependence on the field modes $\phi_{\Vec{k}}$ with $|\Vec{k}|>\mu$.

From a quantum information point of view, Eq. (\ref{effaction}) is exactly the identity $ \langle\mathbb{O}\rangle = \Tr(\rho_A\mathbb{O}_A) = \Tr(\rho\mathbb{O}_A\otimes\mathbb{I}) $ which characterizes completely the reduced density operator $\rho_A$ of a subsystem $A$ \cite{nielsen,entropy}. Thus, the path integral written in terms of the Fourier-transformed fields $\phi_{\Vec{k}}$ reveals that the Hilbert space of a QFT has the tensor product structure $\mathcal{H} = \bigotimes_{\Vec{k}}\mathcal{H}_{\Vec{k}}$ and so entanglement between momentum modes can be characterized.

This means that the Wilsonian effective action $S_\mu[\phi_{|\Vec{k}|\leq\mu}]$ naturally defines a reduced density operator $\rho_\mu$ for momentum modes with $|\Vec{k}|\leq\mu$, with matrix elements given by Eq. (\ref{density}). This allows for the calculation of entanglement measures between scales below and above $\mu$ (as $\rho_\mu$ is the partial trace of a pure state, any entropy is due to entanglement in momentum space). Furthermore, we can conclude that an effective action contains all the required ``information" to define the density matrix associated with a state or subsystem, even when talking about tensor product partitions which are not in momentum space\cite{Balasubramanian:2011wt}.\bigskip

Before moving forward, some comments are in order. First, the Fourier-transformed field considered is labeled by the spatial momentum $\Vec{k}$ without mention of the component associated with the time variable. This is because the actual degrees of freedom in a QFT are spread in space with time indicating their dynamics instead of introducing new variables. Another way of seeing this is through the use of the Euclidean path integral, where the imaginary time and corresponding momentum component are present merely as a trick to projecting states into the vacuum and are thus unrestricted in their corresponding integrals. Second, our focus here is on the ground state for a simple reason: it is well known that all states of a QFT can be generated by linear combinations of local operators acting on the vacuum \cite{haag} and it is the state which determines the thermodynamic phase of the system (at zero temperature, the case here). Thus, studying entanglement in the ground state potentially reveals information about the theory in general.

\subsection{Entanglement measures from the effective action}
\label{measures}
In this subsection we will derive one of the main results of this paper: the construction of a new method for determining entanglement measures associated with the reduced density matrix given by Eq. (\ref{density}).

First, we can modify Eq. (\ref{density}) so that all terms relate only to the low-momentum degrees of freedom (more generally, only to the subsystem variables), as in the current formulation the partition function $Z(\beta)$ in that expression is still the one corresponding to the full system.  When the partial trace is taken, the generated effective action contains a term at zeroth order in the low-momentum fields. By discarding this term, we can define,
\begin{equation}
\label{partition}
    Z(\mu, \beta) := \int_{\beta}\mathcal{D}\phi_{\Vec{k}}(\tau)e^{-S^\beta_\mu[\phi_{\Vec{k}}]},
\end{equation}
and from now on $S_\mu[\phi_{\Vec{k}}]$ is understood as an action not containing any terms independent of the fields. The subindex $\beta$ in the integral sign indicates that integration is taken over paths with time periodicity $\beta$. This in turn adjusts the path integral representation of the matrix elements to,
\begin{equation}
\label{rhomu}
      \bra{\varphi_{\Vec{k}}}\rho_\mu\ket{\Tilde{\varphi}_{\Vec{k}}} = \lim_{\beta\to\infty} \frac{1}{Z(\mu,\beta)} \int_{\phi_{\Vec{k}}(0)=\varphi_{\Vec{k}}}^{\phi_{\Vec{k}}(\beta)=\Tilde{\varphi}_{\Vec{k}}}\mathcal{D}\phi_{\Vec{k}}e^{-S^\beta_\mu}.
\end{equation}
Such a formulation is more practical, being often used implicitly in ordinary renormalization calculations (effective action formalism), where the field-independent free energy term generated by the RG flow is ignored.

Now, with Eq. (\ref{rhomu}) at hand, we can write a formal expression for $\Tr\rho_\mu^n$, where $n$ is an integer. These are themselves valid entanglement measures, generalizations of the so-called purity \cite{nielsen}, and also allow for the calculation of the entanglement entropy via the replica trick. Thus, after performing the matrix multiplication and trace, we obtain, 
\begin{equation}
\label{trace}
    \begin{split}
          \Tr\rho^n_\mu = \lim_{\beta \to \infty} \frac{1}{[Z(\mu,\beta)]^n}\int\mathcal{D}\varphi_1...\int\mathcal{D}\varphi_n  \\
          \times\int_{\varphi_{\Vec{k}}(0)=\varphi_1}^{\varphi_{\Vec{k}}(\beta)=\varphi_2}\mathcal{D}\phi_{\Vec{k}}e^{-S^\beta_\mu}...\int_{\varphi_{\Vec{k}}(0)=\varphi_n}^{\varphi_{\Vec{k}}(\beta)=\varphi_1}\mathcal{D}\phi_{\Vec{k}}e^{-S^\beta_\mu}.
    \end{split}
\end{equation}
By writing the effective action as the integral of an effective Lagrangian, $S^\beta_\mu = \int_0^\beta d\tau L^\beta_\mu$, we perform the following manipulation. In Eq. (\ref{trace}), we shift the limits of integration in $\tau$ of the $(u+1)$th path integral ($u$ being an integer in $\{0,...,n-1\}$) from $[0,\beta]$ to $[u\beta,(u+1)\beta]$. This allows the path integrals to be combined into a single one over fields periodic in $[0,n\beta]$ (it can be seen that due to the trace and operator multiplication, the boundary conditions at multiples of $\beta$ match perfectly and are integrated over).  
These shifts and subsequent recombination are the main reason why a representation of the density matrices employing a finite temperature formalism is used, as at the moment it is not clear how to construct a concrete method directly at zero temperature.

Thus, by combining the effective actions with shifted time variables into a single exponent,
\begin{equation}
\Tr\rho^n_\mu = \lim_{\beta \to \infty} \frac{1}{[Z(\mu,\beta)]^n}\int_{n\beta}\mathcal{D}\varphi_{\Vec{k}}e^{-\sum_u\int_{u\beta}^{(u+1)\beta} d\tau L^\beta_\mu},
\end{equation}
and defining the ``modified partition function",
\begin{equation}
\label{modz}
    Z_n(\mu,\beta):= \int_{n\beta}\mathcal{D}\varphi_{\Vec{k}}(\tau)e^{-\sum_{u=0}^{n-1}\int_{u\beta}^{(u+1)\beta} d\tau L^\beta_\mu},
\end{equation}
the trace can be rewritten as,
\begin{equation}
\label{formula}
    \Tr\rho^n_\mu = \lim_{\beta \to \infty} \frac{Z_n(\mu,\beta)}{[Z(\mu,\beta)]^n}.
\end{equation}
Note the similarities with the expression for real-space entanglement in a QFT given by Eq. (8) of Ref.  \cite{cc}. The difference is that in the momentum-space scenario, there is an effective action allowing the partial trace to be performed, and the high-momenta degrees of freedom are completely ignored. 

It is also important to point out clearly that this path integral method involves a difference in the inverse temperature $\beta$ associated to the degrees of freedom that were traced out, thus defining the effective action and the temperature $n\beta$ for the remaining variables. This difference is key for obtaining the correct results through our method and also appears naturally in other techniques, such as the one derived in \cite{Flynn:2022tbj}.\bigskip

At first glance it might be tempting to assume $\sum_{u=0}^{n-1}\int_{u\beta}^{(u+1)\beta} d\tau L^\beta_\mu = \int_0^{n\beta} L^\beta_\mu$. However, this is not correct. As will be shown in detail in the next Section, the effective action has the general form,
\begin{equation}
    S^\beta_\mu = \int_0^\beta d\tau L_{\rm local}(\tau) + \int_0^\beta d\tau\int_0^\beta d\tau' \Tilde{L}(\tau,\tau')+\dots,
\end{equation}
where $L_{\rm local}(\tau)$ indicates that it is part of a Lagrangian local in time, composed of differential operators $\frac{d}{d\tau}$, while terms like $\Tilde{L}(\tau,\tau')$ involve nonlocal integral kernels. The latter are essential to ensure that the effective action generates a mixed state density matrix (they are also ubiquitous in the study of open quantum systems \cite{open} described by mixed states).

Given such a structure for the effective action, and the fact that any nonlocal terms appear as functions of $\tau-\tau'$, it follows that $\sum_{u=0}^{n-1}\int_{u\beta}^{(u+1)\beta} d\tau L^\beta_\mu$ can be written as,
\begin{equation}
    \label{exponent}
    \begin{split}
        \sum_{u=0}^{n-1}\int_{u\beta}^{(u+1)\beta} d\tau L^\beta_\mu = \int_0^{n\beta} d\tau L_{\rm local}(\tau) \\ +\sum_{u=0}^{n-1}\int_{u\beta}^{(u+1)\beta} d\tau\int_{u\beta}^{(u+1)\beta} d\tau' \Tilde{L}(\tau,\tau')+\dots, 
    \end{split}
\end{equation}
so that the part of the action that is local in time is associated to an integral from $0$ to $n\beta$. On the other hand, the nonlocal one inherits a more complicated structure, which does not simply correspond to a double integral in $[0,n\beta]$. In the next Section it will be shown that this fact leads to a nonzero entropy.

For convenience, we may simplify the notation for the sum of double integrals in Eqs. (\ref{modz}) and (\ref{exponent}) by defining,
\begin{equation}
\label{theta}
\begin{split}
    \Theta_{n}(\tau,\tau'):=\sum_{u=0}^{n-1}\Theta(\tau-u\beta)\Theta(\tau'-u\beta)\times\\
    \Theta[(u+1)\beta-\tau]\Theta[(u+1)\beta-\tau'],
\end{split}
\end{equation}
where $\Theta(\tau)$ is the step function. With this, we have, 
\begin{equation}
    \label{simple}
    \begin{split}
        \sum_{u=0}^{n-1}\int_{u\beta}^{(u+1)\beta} d\tau\int_{u\beta}^{(u+1)\beta} d\tau' \Tilde{L}(\tau-\tau')\\
        =\int_0^{n\beta}d\tau\int_0^{n\beta}d\tau'\Theta_{n}(\tau,\tau')\Tilde{L}(\tau-\tau').
    \end{split}
\end{equation}
 In this form it is also easy to see how to generalize the expressions in case the effective action involves integrals over three or more time variables.\bigskip

Finally, once $\Tr\rho_\mu^n$ is obtained, the Rényi entropies are given by,
\begin{equation}
    H_n(\mu) = \frac{1}{n-1} \lim_{\beta \to \infty} \left( n\log Z(\mu,\beta)-\log Z_n(\mu,\beta)\right),
\end{equation}
and, as usual, the entanglement entropy is derived through the formal limit $S_{EE}(\rho_\mu) = \lim_{n\to 1} H_n(\mu)$, meaning that calculating $Z_n(\mu,\beta)$ is the key step in deriving entanglement measures from an effective action.\bigskip

To show that nonlocal terms are indeed crucial in obtaining the entropy, consider that under some approximation scheme the effective action $S_\mu^\beta$ is taken to contain only local terms in time, that is, $ S^\beta_\mu = \int_0^\beta d\tau L_{local}(\tau)$. Via the Legendre transform, we may obtain an associated Hamiltonian $\Tilde{H}$ with ``thermal partition function" given (via the usual path integral construction) by,
\begin{equation}
    \Tr e^{-\beta \Tilde{H}} = \int_{\beta} \mathcal{D}\phi_{\Vec{k}} e^{-\int_0^\beta d\tau L_{local}(\tau)},
\end{equation}
and, therefore, $\Tr\rho_\mu^n$ is,
\begin{equation}
   \begin{split}
        \Tr\rho^n_\mu &= \lim_{\beta \to \infty} \frac{\int_{n\beta} \mathcal{D}\phi_{\Vec{k}} e^{-\int_0^{n\beta} d\tau L_{local}(\tau)}}{\left[\int_{\beta} \mathcal{D}\phi_{\Vec{k}} e^{-\int_0^\beta d\tau L_{local}(\tau)}\right]^n} \\
        &= \lim_{\beta \to \infty} \frac{\Tr e^{-n\beta \Tilde{H}}}{(\Tr e^{-\beta \Tilde{H}})^n} = 1.
   \end{split}
\end{equation}
The last equality comes from diagonalizing $\Tilde{H}$ to calculate the traces. It is basically the well-known statement that thermal states of a Hamiltonian approach the vacuum, a pure state with no entropy, as the temperature goes to zero. With this, we see that, as claimed, the nonlocal (in Euclidean time) terms of the effective action are essential for the entanglement entropy not to vanish.

\textcolor{black}{The role of this Euclidean time non-locality in describing mixed states (and thus entanglement in our case) has been studied in Ref. \cite{Agon:2014uxa} in the operator formalism as a consequence of the non-Hamiltonian evolution of open quantum systems. There a similar conclusion is reached by deriving the Kraus operators for the time evolution of the low-momentum degrees of freedom under perturbation theory and certain conditions.}

\section{Applications of the method}
\label{applications}

With the method developed in the previous Section, we can in principle calculate the entropies associated with any density operator $\rho$ whose matrix elements are generated by a path integral of some effective action $S_{eff}$ in Euclidean time, with the corresponding calculation being roughly that of a partition function. Note that this technique is applicable even if the entropy of $\rho$ is not associated with momentum-space variables or is not due to entanglement at all. Thus, in this Section, we will first calculate the entanglement entropy of two coupled quantum harmonic oscillators in perturbation theory. Since this entropy has already been found by other means \cite{nishioka}, this calculation offers a benchmark for checking the validity of the method. We will then move to the main topic of interest in this paper and calculate entanglement measures in momentum space of QFTs where the used low-momentum effective action is obtained via the Wilsonian procedure of integrating out fast modes. The theories studied in this paper feature real scalar fields with $\phi^3$ and $\phi^4$ interactions in the perturbative regime, and the momentum space entropies will be calculated only up to the lowest order in the coupling, which already leads to a non-zero result. 

\subsection{Coupled Harmonic Oscillators}
\label{harmonic}

The review in Ref. \cite{nishioka} considers a quantum system with two particles, with positions denoted $x_A$ and $x_B$, moving in one dimension inside a quadratic potential and linearly coupled to each other, and calculates exactly the entanglement entropy between the particles in the ground state. This is done by taking the wavefunction of this state, tracing over $x_B$ and diagonalizing exactly the resulting reduced density matrix $\rho_A$. 

Thus, having a known result to compare to, we now apply our method to the ground state of this system. The Euclidean Lagrangian of the model is given by, 
\begin{equation}
    L= \frac{1}{2}\left(-\frac{d^2}{d\tau^2}+M^2\right)\left(x_A(\tau)^2+x_B(\tau)^2\right) -lx_Ax_B,
\end{equation}
where $M^2$ can be related to the parameters used in Ref. \cite{nishioka}.

Applying the technique consists of performing the path integral over $x_B$ to generate an effective action for $x_A$ alone (discarding any terms independent of $x_A$ which may appear) and, from this effective action, calculate at finite temperature the associated $Z_n(A, \beta)$ and $Z(A,\beta)$.

The path integral over $x_B$ can be easily performed by going to frequency space and it leads to the effective action,
\begin{equation}
    S_{\rm eff} = \frac{1}{2}\int \frac{d\omega}{2\pi} \left(\omega^2+M^2- \frac{l^2}{\omega^2+M^2}\right)|x_A(\omega)|^2.
\end{equation}
Returning to imaginary time, $S_{eff}$ becomes,
\begin{equation}
    S_{\rm eff} = \frac{1}{2}\int d\tau d\tau'x_A(\tau)A(\tau,\tau')x_A(\tau'), 
\end{equation}
where, 
\begin{equation}
    A(\tau,\tau') = \left(-\frac{d^2}{d\tau^2}+M^2\right)\delta(\tau-\tau')-\frac{l^2}{2M}e^{-M|\tau-\tau'|},
\end{equation}
with, as argued previously, a term exhibiting nonlocality in time appearing in the effective action. 

Since all calculations must be done at finite temperature, the nonlocal kernel is actually,
\begin{equation}
\label{sum}
    \frac{1}{\beta}\sum_n\frac{e^{-i\omega_n(\tau-\tau')}}{\omega_n^2+M^2} = \frac{e^{-M|\tau-\tau'|}}{2M}+\frac{\cosh(M|\tau-\tau'|)}{M(e^{\beta M}-1)}.
\end{equation}
As shown in Appendix \ref{a1}, due to the $e^{\beta M}$ factor in the denominator, the second term goes to zero exponentially as $\beta\to\infty$, thus not affecting the zero-temperature entropy. Hence, it can be ignored in this calculation (this is not the case when the effective action is non-gaussian, as we will see in the next subsections).

In order to calculate $Z(A,\beta)$, we use the finite temperature effective action given by,
\begin{equation}
    S_{eff}[x_j] = \frac{1}{2}\sum_j \left(\omega_j^2+M^2-\frac{l^2}{\omega_j^2+M^2}\right)x_j^*x_j,
\end{equation}
with $\omega_j = \frac{2\pi j}{\beta}$ the Matsubara frequencies.
The Gaussian path integral over all $x_j$ is straightforward and leads to the expression, 
\begin{equation}
    \log Z(A,\beta) = -\frac{1}{2}\sum_j\log\left(\omega_j^2+M^2-\frac{l^2}{\omega_j^2+M^2}\right),
\end{equation}
which, after employing simple algebraic manipulations and known Matsubara sums, becomes, 
\begin{equation}
    Z(A,\beta) = \frac{\sinh(\frac{\beta M}{2})}{2\sinh(\frac{\beta\sqrt{M^2+l}}{2})\sinh(\frac{\beta\sqrt{M^2-l}}{2})}. 
\end{equation}

For the next step, which is calculating $Z_n(A, \beta)$, it is necessary to perform the particular sum $\sum_{u=0}^{n-1}\int_{u\beta}^{(u+1)\beta} d\tau L^\beta_\mu$. As shown previously, the local terms simply add up to an ordinary integral from $0$ to $n\beta$, so the focus now is on,
\begin{equation}
    \int_0^{n\beta} d\tau\int_0^{n\beta} d\tau'\Theta_{n}(\tau,\tau') e^{-M|\tau-\tau'|}x_A(\tau)x_A(\tau').
\end{equation}
The calculations from this point on are quite extensive and the details are relegated to Appendix \ref{a1}. Ultimately, up to order $\mathcal{O}(l^2)$ in the perturbative regime, we find,
\begin{equation}
    \Tr\rho_A^n = 1-n\frac{l^2}{16M^4}.
\end{equation}
In Ref. \cite{nishioka} this trace is calculated exactly and is given by $\Tr\rho_A^n = \frac{(1-\xi)^n}{1-\xi^n}$, with $\xi = \left(\frac{(M^2+l)^{\frac{1}{4}}-(M^2-l)^{\frac{1}{4}}}{(M^2+l)^{\frac{1}{4}}+(M^2+l)^{\frac{1}{4}}}\right)^2$ in our notation. Expanding the exact result up to order $l^2$, the same result is obtained.

\subsection{Perturbative calculation in $\phi^3$ theory}
\label{phi3}

In this and the next subsections, we will calculate the entanglement between the degrees of freedom at different momentum scales of perturbative scalar theories (as always, in the ground state). From the discussion in the previous Section, this entanglement will be directly related to the Wilsonian effective action and also be given a diagrammatic interpretation.

The first step of the calculation is splitting the field variable as a sum of high and low momentum parts (the separation being determined by a chosen scale $\mu$) and integrate the high momentum modes perturbatively, introducing at first an overall UV cutoff $\Lambda$, in order to find the effective action $S_\mu[\phi_{|\Vec{k}|\leq\mu}]$. The chosen order of perturbation theory will the lowest one in which a nonlocal term in time appears. After this, we apply the method constructed earlier to calculate the nth order Rényi entropies. 

Integrating out modes with spatial momentum obeying $|\Vec{k}|>\mu$, the perturbative corrections to the effective action are obtained by the usual connected Feynman diagrams under the condition that all internal lines have momentum above the scale $\mu$ \cite{wilson}.

For the $\phi^3$ theory in spacetime dimension $d$, we begin with the Euclidean bare action,
\begin{equation}
    S[\phi] = \int d^dx\left[\frac{1}{2}(\nabla\phi)^2+\frac{1}{2}m^2\phi^2+\frac{\lambda}{3!}\phi^3\right].
\end{equation}

At order $\lambda$ the only contribution to $S_\mu$ besides the already existing $\lambda\phi^3$ comes from the ``tadpole" diagram. However, this only shifts the expectation value of $\phi$ and is local in $\tau$, so in view of the discussion in Section \ref{lowmomentum} this yields $\Tr\rho_\mu^n = 1+\mathcal{O}(\lambda^2)$, and the first order generation of entanglement is zero.

Now we will begin to use Feynman diagrams in earnest and since we are performing the Wilsonian integration of fast modes, we will use solid lines to denote momenta $\Vec{k}$ such that $|\Vec{k}|<\mu$,  and dashed lines for $|\Vec{k}|>\mu$. As usual, all internal lines must be dashed while all external ones must be solid.

In the $\phi^3$ theory, the diagrams with two vertices for the effective action have the form given by, 
\begin{figure}[H]
	\centering
	\includegraphics[width=3cm]{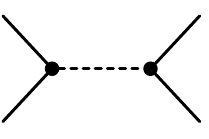}
\end{figure}
%
\noindent
corresponding to a $\phi^4$ term in the effective action given at finite temperature by, 
\begin{equation}
\label{4phi3}
   \frac{\lambda^2}{8} \frac{\phi_{j_1,\Vec{k}_1}\phi_{j_2,\Vec{k}_2}\phi_{j_3,\Vec{k}_3}\phi_{-j_1-j_2-j_3,-\Vec{k}_1-\Vec{k}_2-\Vec{k}_3}}{(\omega_{j_1}+\omega_{j_2})^2+(\Vec{k}_1+\Vec{k}_2)^2+m^2}
\end{equation}
We also get the one-loop diagram,
\begin{figure}[H]
	\centering
	\includegraphics[width=3cm]{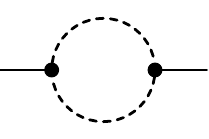}
\end{figure}
%
\noindent
translating to,
\begin{equation}
\label{massphi3}
   \frac{1}{2}\times\frac{\lambda^2}{2}
        \frac{1}{\omega_{j'}^2+\Vec{q}^2+m^2}\frac{\phi_{j,\Vec{k}}\phi_{-j,-\Vec{k}}}{(\omega_j+\omega_{j'})^2+(\Vec{k}-\Vec{q})^2+m^2}
\end{equation}
Here attention has to paid to the extra factor $1/2$: it arises because it is not the mass renormalization per se that is being calculated, which would eliminate this factor in view of the structure of the Lagrangian. Rather, we are working with the full numerical factor of the Feynman diagram.

The effective action includes the sums over Matsubara frequencies (with the field normalized as $\phi(\tau) = \frac{1}{\sqrt{\beta}}\sum_je^{i\omega_j\tau}\phi_j$ to give the correct number of $\beta^{-1}$ factors) and integrals over appropriate momentum regions as defined by the rules of the Wilson RG.

Taking the Fourier transform of the time component of the fields, these diagrams indeed lead to nonlocal terms. For illustration we show the respective zero-temperature kernels, as those are simpler (the finite-temperature ones we need to actually use in our method are discussed in Appendix \ref{a2}),
\begin{equation}
\label{nonlocalextra}
  \frac{e^{-\sqrt{(\Vec{k}_1+\Vec{k}_2)^2+m^2}|\tau-\tau'|}}{2\sqrt{(\Vec{k}_1+\Vec{k}_2)^2+m^2}}\phi_{\Vec{k}_1}(\tau)\phi_{\Vec{k}_2}(\tau)\phi_{\Vec{k}_3}(\tau')\phi_{\Vec{k}_4}(\tau'),
\end{equation}
for the generated $\phi^4$ term, and, 
\begin{equation}
\label{nonlocalphi3}
  \frac{e^{-(\sqrt{\Vec{q}^2+m^2}+\sqrt{(\Vec{k}-\Vec{q})^2+m^2})|\tau-\tau'|}}{4\sqrt{\Vec{q}^2+m^2}\sqrt{(\Vec{k}-\Vec{q})^2+m^2}}\phi_{\Vec{k}}(\tau)\phi_{-\Vec{k}}(\tau'),
\end{equation}
for the correction to the $\phi^2$ term.

In both cases (and in general) the nonlocality appears because at least one external momentum $\Vec{k}$ appears in one of the propagators, thus leading to the above kernels when the Fourier transforms are performed. This is the reason why tadpoles like the $\mathcal{O}(l)$ diagram are local in time.

The exponential structure of the nonlocal kernel of Eqs. (\ref{nonlocalextra}) and (\ref{nonlocalphi3}) allows many of the calculations made for the system of coupled harmonic oscillators to be adapted to this case. More generally, this is a direct consequence of perturbation theory, since diagrams generate products and convolutions of propagators, whose Fourier transforms, before performing the spatial momentum integrals, are exponential functions.

Now, referring to Appendix \ref{a2} for details of the main calculation, we calculate the logarithm of the modified partition function $\log Z_n(\mu,\beta)$ (as usual for field theories, the logarithm is more practical) following the same strategy as in the previous case: by expanding the exponential of the action up to $\mathcal{O}(\lambda^2)$ and using Wick's theorem on the products of fields which appear, the nonlocal terms are averaged over the original free action at temperature $n\beta$ and their contributions are summed. In particular, this means that each term of the Rényi entropy can be interpreted as coming from the \textit{connected vacuum bubbles derived from the nonlocal diagrams via contractions of their free legs}. This interpretation is possible because the field contractions via the Wick theorem are represented by connecting the free legs of the diagrams associated with nonlocal terms. Therefore, for the two-legged diagram in $\phi^3$ theory we have,
\begin{figure}[H]
	\centering
	\includegraphics[width=4cm]{graph2p-dashed-loop.1.pdf}

\scalebox{2}{$\Downarrow$}

\includegraphics[width=4cm]{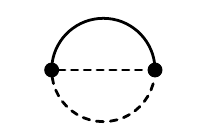}
\end{figure}
%
%
%
\noindent
while the diagram with four legs has the possible contractions,
\begin{figure}[H]
	\centering
	\includegraphics[width=4cm]{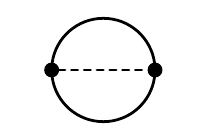}
	
	and
	
	\includegraphics[width=4cm]{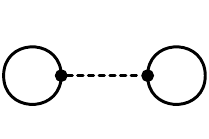}
\end{figure}
In this last case, the difference in the structures of the bubbles is very important. The internal line on the second diagram above must have, by definition, spatial momentum with magnitude greater than $\mu$.  However, momentum conservation imposed at the vertices forces it to vanish, and the impossibility of fulfilling both conditions at the same time implies that this diagram automatically vanishes. Hence, only the ``basketball" diagram contributes to the entropy. We will see that a similar behavior also occurs in the $\phi^4$ case.

At the end of all calculations we find that the Wick contractions generate delta functions for the spatial momenta such that the $n$th Rényi entropy will always be an extensive quantity, a result also obtained in Ref. \cite{Balasubramanian:2011wt} (this makes sense, as the momentum degrees of freedom are uniformly spread throughout space and so the total entropy should be proportional to the volume of the system). Furthermore, the entropy \textit{density} resulting from the sum of terms associated with the bubble diagrams discussed above is,
\begin{equation}
    \label{entropyphi3}
    \frac{ H_n(\mu)}{V} = \frac{n}{n-1} \frac{\lambda^2}{8}\int^*\frac{d^{d-1}k}{(2\pi)^{d-1}}\frac{d^{d-1}q}{(2\pi)^{d-1}} \mathcal{I}(\Vec{k}, \Vec{q}),
\end{equation}
with,
\begin{equation}
  \begin{split}
      \mathcal{I}(\Vec{k}, \Vec{q})\equiv  \frac{1}{\sqrt{\Vec{k}^2+m^2}}\frac{1}{\sqrt{\Vec{q}^2+m^2}}
    \frac{1}{\sqrt{(\Vec{k}-\Vec{q})^2+m^2}}\\
    \times
     \frac{1}{\left(\sqrt{\Vec{k}^2+m^2}+\sqrt{\Vec{q}^2+m^2}+\sqrt{(\Vec{k}-\Vec{q})^2+m^2}\right)^2},
  \end{split}
\end{equation}
\noindent
where the spatial components of the momenta are integrated over the region such that, given $\Vec{k}, \Vec{q}$ and $\Vec{k}-\Vec{q}$, at least one of them is below the scale $\mu$, at least one is above it and no set of momenta is repeated in the integration. Such a specific region is a direct consequence of the structure of basketball Feynman diagrams: as mentioned earlier, the number of solid and dashed lines in each bubble indicates how many momenta are integrated over magnitudes smaller and greater than $\mu$, respectively. Furthermore, the repetition of lines of a same type in a diagram means we can multiply the associated expression by a symmetry factor at the cost of forbidding repeating sets of momenta in the integration region.

The reason for this specific manipulation is that it simplifies the final analytical expression and allows us to compare directly Eq. (\ref{entropyphi3}) to the results from Ref. \cite{Balasubramanian:2011wt}.\bigskip

Before moving on, note that while the $\phi^3$ theory is obviously problematic as the energy is not bounded from below, the perturbative result we find above is actually associated with a $\phi^3$ vertex in any theory containing such term in the Lagrangian, and so it is still of value in the actual physically relevant model.

A more important point is that our result reproduces exactly the one from Ref. \cite{Balasubramanian:2011wt} (for the lowest order Rényi entropy which can be derived through their method), but it was now obtained directly employing the Wilsonian point of view, and it also gives a diagrammatic interpretation that arises naturally from the calculation. Furthermore, we can in addition postulate the following ``Feynman rules for Rényi entropy":

\begin{equation}
\label{Eq:Prop}
\vcenter{\hbox{\includegraphics[width=2cm]{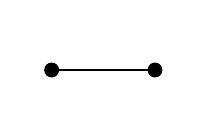}}}=
\frac{1}{2\sqrt{\Vec{k}^2+m^2}} = \int_{-\infty}^{\infty}\frac{dk_0}{2\pi}G_0(k_0,\Vec{k}),
\end{equation}

\begin{equation}
\vcenter{\hbox{\includegraphics[width=2cm]{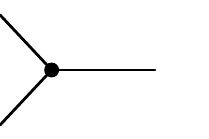}}}=
\frac{\lambda(2\pi)^{d-1}\delta^{d-1}(\Vec{k}+\Vec{p}+\Vec{q})}{\sqrt{\Vec{k}^2+m^2}+\sqrt{\Vec{p}^2+m^2}+\sqrt{\Vec{q}^2+m^2}},
\end{equation}
with solid or dashed lines, depending on whether they represent slow or fast modes.\bigskip

By applying these rules to the bubble diagrams shown previously, integrating momenta over the specific region discussed and including $n/(n-1)$ as a prefactor, the lowest order result is reproduced correctly. Note that by the rules given every diagram will produce a factor of $(2\pi)^{d-1}\delta^{d-^1}(0)$, which becomes the total volume when defining the theory in a finite box, thus we have extensive entropies as expected from physical intuition and the direct calculation done in the Appendices.\bigskip

So far only the Rényi entropies were discussed. However, the entanglement entropy at lowest order is proportional to them. As a consequence, all conclusions in this subsection apply to that entanglement measure as well. This will be proven in subsection \ref{perturb}, but first we will do a similar study for the case of a $\phi^4$ interaction.

\subsection{Results for $\phi^4$ theory}
\label{phi4}

Drawing from the lessons of the previous sections, we can now proceed and calculate the momentum-space entanglement for $\lambda\phi^4$ theory. This QFT has as the bare action:

\begin{equation}
   S[\phi] = \int d^dx\left[\frac{1}{2}(\nabla\phi)^2+\frac{1}{2}m^2\phi^2+\frac{\lambda}{4!}\phi^4\right]
\end{equation}

Integrating out modes with momentum $\Vec{k}$ such that $|\Vec{k}|>\mu$ perturbatively, the discussion of Section \ref{phi3} made clear that in order to get a finite entropy only diagrams which have external momenta in an internal line contribute at lowest order.

At order $\lambda$ the only Feynman diagram is the tadpole, which yields a mass renormalization without generating a nonlocal term in time. At order $\lambda^2$ three diagrams will lead to nonlocality in time: 
%
%
%
\begin{figure}[H]
	\centering
	\includegraphics[width=3cm]{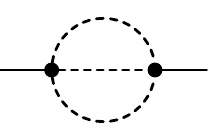}
	
	\includegraphics[width=3cm]{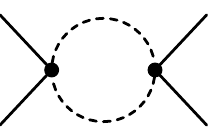}
	
	\includegraphics[width=3cm]{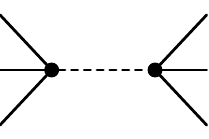}
\end{figure}
\noindent
These lead to the respective terms (written at zero-temperature for simplicity),
\begin{equation}
\label{twofield}
    \frac{1}{2}\times\frac{\lambda^2}{6}\frac{1}{q^2+m^2}\frac{1}{p^2+m^2}\frac{1}{(k-q-p)^2+m^2}\phi_k^*\phi_k,
\end{equation}
\begin{equation}
\label{fourfield}
    \frac{\lambda^2}{16}\frac{1}{q^2+m^2}\frac{(2\pi)^d\delta(\sum_ik_i)}{(k_1+k_2-q)^2+m^2}\phi_{k_1}\phi_{k_2}\phi_{k_3}\phi_{k_4},
\end{equation}
\begin{equation}
\label{sixfield}
    \frac{\lambda^2}{72}\frac{(2\pi)^d\delta(\sum_{i=1}^6k_i)}{(k_1+k_2+k_3)^2+m^2}\phi_{k_1}\phi_{k_2}\phi_{k_3}\phi_{k_4}\phi_{k_5}\phi_{k_6},
\end{equation}
with the inclusion of integrals over specific momentum regions arising from tracing out high-momentum modes. Once again we have an extra $1/2$ factor in the two-point contribution like the one discussed for the $\phi^3$ theory. This can also be seen directly in Eqs. (4.9) and (4.20) of Ref. \cite{wilson}

The next step is to find how exactly these new expressions are nonlocal in imaginary time. Following Appendix \ref{a3}, we have exponential kernels of the form (suppressing some terms for simplicity),
\begin{equation}
 \frac{e^{-(\sqrt{\Vec{q}^2+m^2}+\sqrt{\Vec{p}^2+m^2}+\sqrt{(\Vec{k}-\Vec{q}-\Vec{p})^2+m^2})|\tau-\tau'|}}{8\sqrt{\Vec{q}^2+m^2}\sqrt{\Vec{p}^2+m^2}\sqrt{(\Vec{k}-\Vec{q}-\Vec{p})^2+m^2}},
\end{equation}
\begin{equation}
   \frac{e^{-(\sqrt{\Vec{q}^2+m^2}+\sqrt{(\Vec{k}_1+\Vec{k}_2-\Vec{q})^2+m^2})|\tau-\tau'|}}{4\sqrt{\Vec{q}^2+m^2}\sqrt{(\Vec{k}_1+\Vec{k}_2-\Vec{q})^2+m^2}},
\end{equation}
\begin{equation}
   \frac{e^{-\sqrt{(\Vec{k}_1+\Vec{k}_2+\Vec{k}_3)^2+m^2}|\tau-\tau'|}}{2\sqrt{(\Vec{k}_1+\Vec{k}_2+\Vec{k}_3)^2+m^2}}
\end{equation}

Similarly to the $\phi^3$ theory, in the lowest order calculation each of the terms above will contribute to the entropy independently of the other, so that the end result will simply be their sum. Furthermore, the structure of the end results can be associated with the possible vacuum bubbles obtained by contracting the legs of the diagrams, namely, 
%
\begin{figure}[H]
	\centering
	\includegraphics[width=3cm]{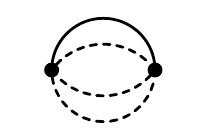}
\end{figure}
\noindent
from the ``sunrise" diagram, and,  
\begin{figure}[H]
	\centering
	\includegraphics[width=3cm]{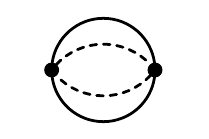}
	
	and
	
	\includegraphics[width=4cm]{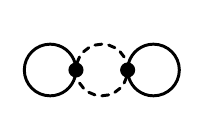}
\end{figure}

\noindent
from the one-loop correction of the coupling, along with,  
%
%
\begin{figure}[H]
	\centering
	\includegraphics[width=3cm]{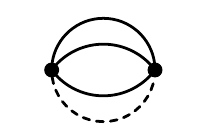}
	
	and
	
	\includegraphics[width=4cm]{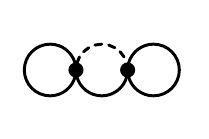}
\end{figure}
\noindent
from the new $\phi^6$ term.\bigskip

In the last two cases we are faced again with both ``basketball" and ``cactus" vacuum bubbles and here, too, we find that the latter type of diagram is canceled and does not contribute to the entropy. 

The cancellation for the cactus diagram associated with the four-field term is explained in Appendix \ref{a3} along with the remaining steps of the overall calculation. As for the cactus diagram arising from $\phi^6$, the reason for it to vanish is simple: momentum conservation and the structure of the diagram imply that each individual loop must have the same value of momentum throughout its extension. However, in this case the middle loop is a half solid and a half dashed line, while no momentum can be a fast and slow mode at the same time, so there will be at least one unsatisfied delta function and this sends the whole expression to zero.   

After all the calculations, we again use the permutation symmetry of lines of the same type in each vacuum diagram to combine all contributions into a single analytical expression (with an intricate momentum integration region as before), leading to the $n$th Rényi entropy density:
\begin{equation}
\label{entropyphi4}
    \frac{H_n(\mu)}{V} = \frac{n}{n-1} \frac{\lambda^2}{16}\int^*\prod_{i=1}^3\frac{d^{d-1}k_i}{(2\pi)^{d-1}}\mathcal{I}(\Vec{k}_1,\Vec{k}_2,\Vec{k}_3),
\end{equation}
with,
\begin{equation}
\begin{split}
\mathcal{I}(\Vec{k}_1,\Vec{k}_2,\Vec{k}_3) \equiv \frac{1}{\sqrt{(\sum_{i=1}^3\Vec{k}_i)^2+m^2}}\prod_{i=1}^3\frac{1}{\sqrt{\Vec{k}_i^2+m^2}}\\
 \times \frac{1}{\left(\sum_{i=1}^3\sqrt{\Vec{k}_i^2+m^2}+\sqrt{(\sum_{i=1}^3\Vec{k}_i)^2+m^2}\right)^2}
    \end{split},
\end{equation}
and where the integration limits, like in the $\phi^3$ case, are such that at least one momentum is below $\mu$, at least one is above it, and no set of momenta is repeated.\bigskip

We recover once more the same Rényi entropy obtained from the method employed in Ref. \cite{Balasubramanian:2011wt} and, once again, the expression could also be obtained by postulating Feynman rules with the propagator line defined as in Eq. (\ref{Eq:Prop}), along with the vertex,
%
%
\begin{equation}
\vcenter{\hbox{\includegraphics[width=2cm]{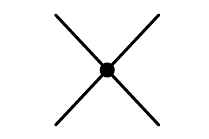}}}=
\frac{\lambda(2\pi)^{d-1}\delta^{d-1}(\sum_{i=1}^4\Vec{k}_i)}{\sum_{i=1}^4\sqrt{\Vec{k}_i^2+m^2}},
\end{equation}
which lead to the correct result by incorporating the prefactor $\frac{n}{n-1}$ and integrating over the specific set of momenta as discussed (and, again, we gain an overall volume factor from the extra delta functions present in each diagram).

\subsection{Perturbation theory and the replica trick}
\label{perturb}
Now, as mentioned in Subsection \ref{a2}, we only derived expressions for the Rényi entropies and avoided references to the $n\to1$ analytical continuation of the replica trick, which gives the entanglement entropy per se. In order to discuss this in detail, let us consider the general form of the $n$th Rényi entropies found throughout this Section,
\begin{equation}
    H_n(\rho_A) = \frac{1}{1-n}\log\Tr\rho_A^n =\frac{n}{n-1}\lambda^2C+\mathcal{O}(\lambda^3),
\end{equation}
where $C$ is some theory-dependent expression.

Clearly, by taking the limit $n\to1$ naively we would arrive at the absurd conclusion that $S_{A}=\infty$. The reason for this is that the expansion in the parameter $\lambda$ is made at fixed $n$ and terms like $\lambda^{2n}$ are ignored for being of higher-order than desired. This means that terms which are important for the entanglement in the limit $n\to1$ are thrown away and cannot be recovered via the limiting procedure. Nevertheless, there is a way of finding the lowest-order contribution to the entanglement entropy through this method. Assume that the reduced density matrix $\rho_A$ is diagonalized exactly and its eigenvalues (as functions of $\lambda$) are given by $p_i(\lambda)$. For $\lambda=0$ all but one of the probabilities must be zero, since the starting point of the perturbative expansion made here is a separable state and, as seen throughout this Section, the entanglement is generated only at order $\lambda^2$ and above. Therefore, these probabilities may be labeled such that $ p_i(\lambda)=\lambda^2a_i(\lambda)$ for $i\geq 1$ and $p_0 = 1-\lambda^2\sum_ia_i(\lambda)$.
Thus, calculating the entanglement entropy from these probabilities:
\begin{equation}
      \begin{split}
           S_A &= -\lambda^2\sum_{i=1}^\infty a_i(\lambda)\log(\lambda^2a_i(\lambda))\\
           &-\left(1-\lambda^2\sum_{i=1}^\infty a_i(\lambda)\right)\log\left(1-\lambda^2\sum_{i=1}^\infty a_i(\lambda)\right),
      \end{split}
\end{equation}
and taking the dominant term as $\lambda\to0$, we find,
\begin{equation}
     S_A = -\sum_{i=0}^\infty p_i\log p_i = -\lambda^2\log \lambda^2 \sum_i^\infty a_i(0) +\mathcal{O}(\lambda^2).
\end{equation}
The presence of a term $-\lambda^2\log \lambda^2$ is ubiquitous in the perturbative regime, see Ref. \cite{Balasubramanian:2011wt} and the exact result in Ref. \cite{nishioka}, and is particular to the entanglement entropy, as the $x\log x$ function is non-analytical at $x=0$.\bigskip

Following the same procedure to calculate any Rényi entropy \textcolor{black}{(with a similar discussion made in Appendix C of Ref. \cite{Agon:2014uxa})} leads to,
\begin{equation}
       H_n(\rho_A)= \frac{1}{1-n}\log\left(\sum_{i=0}^\infty p_i^n\right) \approx \frac{n\lambda^2}{n-1}\sum_i^\infty a_i(0)
\end{equation}
This means that we may find the entanglement entropy by making the substitution $\frac{n}{n-1}\to\log\frac{1}{\lambda^2}$ and the entropies are really proportional to each other at this first approximation (keeping in mind that terms of order $\lambda^2$ or higher are being discarded and that these must be calculated through more sophisticated procedures). Finally, we point out that, strictly speaking, the parameter appearing inside the logarithms must be the square of the \textit{adimensional} coupling constant $\tilde{\lambda}$ of the coupling constant, currently this distinction does not affect the results or their interpretation in any relevant way but it is important to keep it in mind when proceeding to higher orders of the perturbative expansion.

\section{Conclusions and Outlook}
\label{conclusion}

We have developed a path integral method to compute the entanglement between high and low momentum scales which is based on the Wilson RG, where fast momentum modes are integrated out to obtain an effective theory. As discussed previously in the literature \cite{Balasubramanian:2011wt}, the Wilson RG naturally provides a framework where different scales are entangled, since defining fast and slow modes necessarily imply that a partition in momentum space has to be made. We have shown here that strictly adhering to Wilson's prescription using a path integral formalism, one can systematically compute the Rényi entropies, in particular to any order in perturbation theory. This can be done in a simpler fashion than with other methods employed in the literature, since cumbersome matrix diagonalizations are not needed, though the limiting procedure of the replica trick to obtain the entanglement entropy must be handled with care. One reason why this method is appealing and efficient relies on the fact that a Feynman diagram technique can be implemented to facilitate the task: the structure of contractions in the Wilsonian effective action and partition functions is the same as in the usual calculations. However, as far as the Feynman \textit{rules} are concerned, we have only explicitly shown examples at lowest non-trivial order. It remains to show that Feynman rules for the entropies apply equally well at any order of perturbation theory. In a related vein, it is worth mentioning that an extension of the method to study QFTs in the nonperturbative regime is also possible, like for example the $1/N$ expansion, where Feynman diagrams occur in dressed form, thus accounting for an infinite number of diagrams to be resummed using $1/N$ as control parameter rather than the coupling constant.      

There are several other avenues to explore using the method described in detail here. We have only given examples of calculations for scalar field theories, but the method should of course applies equally well to theories involving fermions. 
\textcolor{black}{ 
	However, the application of the method to gauge theories raises a number of questions we intend to explore in a further work. The well known fact that path integrals for gauge fields include redundant degrees of freedom that have to be carefully accounted for may be a source of complications in the implementation. Furthermore, there is also a difficulty related to the Wilson RG itself, whose separation of fast and slow modes breaks gauge invariance at intermediate steps of the calculation.  A way forward could be connecting
methods such as those in Refs. \cite{REUTER1994181,liao1,liao2}, which are gauge-invariant by construction, to the low-momentum reduced density matrix and from this relation deriving a formula for the entropy.}

\textcolor{black}{Conceptually, entanglement of gauge degrees of freedom differs from that of other theories even in real space, where it is currently understood that edge modes must be considered when studying the entanglement between a region of space and its complement in order to obtain sensible results \cite{Buividovich:2008yv,Donnelly:2011hn,Ghosh:2015iwa,Casini:2013rba}. This is because even when regularizing the theory in a lattice, the physical Hilbert space does not factorize as a tensor product labeled by spatial regions; the real lattice gauge theory degrees of freedom are Wilson loops, as discussed in Ref. \cite{Buividovich:2008yv}.  Thus, moving to momentum-space entanglement we can question whether the physical Hilbert space of the theory still factorizes in momentum space and which degrees of freedom are involved in case the factorization occurs (for instance, do edge modes also arise in this case?). These are interesting subtleties we intend to study in a future work.}

Beyond the practical advantages of the technique developed in this paper, there are also fundamental questions that immediately come into focus. For instance, it would be important to investigate the precise meaning of the entanglement between RG scales regarding the fixed point structure of the theory. Does it reveal something deeper about entanglement in QFTs and scale invariance? More precisely, is it possible to have entanglement between momentum scales in a scale-invariant theory (e.g., a theory at its IR fixed point)? Answering such a question would be of paramount importance for quantum information aspects of QFTs. 

Finally, the method developed here also applies to the study of entanglement in open quantum systems or between different types of fields, say bosons and fermions in the Yukawa theory. The technique only requires that the effective action after integrating out some variables is nonlocal in time, so there is a priori no reason to restrict it just to momentum modes.

\begin{acknowledgments}
We thank the Deutsche Forschungsgemeinschaft (DFG) 
for support through the W\"urzburg-Dresden Cluster of Excellence on Complexity and Topology in Quantum Matter – ct.qmat (EXC 2147, project-id 39085490) and 
the Collaborative Research Center  
SFB 1143 (project-id 247310070). M.H.M.C. was supported by Coordena\c{c}\~ ao Aperfei\c{c}oamento de Pessoal de N\'{i}vel Superior - CAPES, grant no. 88887.374238/2019-00. G.K was supported in part by Conselho Nacional de Desenvolvimento Cient\'{i}fico e Tecnol\' ogico - CNPq, grant no. 309262/2019-4 and Funda\c{c}\~ ao de Amparo \`a Pesquisa do Estado de S\~ ao Paulo (FAPESP), grant No. 2018/25225-9.
\end{acknowledgments}

\appendix
\section{Entropy of coupled harmonic oscillators}
\label{a1}

After tracing/integrating out one of the oscillators in Section \ref{harmonic}, the effective action of the remaining degree of freedom is given by,
\begin{equation}
    S_{eff} = \frac{1}{2}\int_0^{\beta}\int_0^{\beta} d\tau d\tau'x_A(\tau)A(\tau,\tau')x_A(\tau'),
\end{equation}
where,
\begin{equation}
    \begin{split}
        A(\tau,\tau') = \left(-\frac{d^2}{d\tau^2}+M^2\right)\delta(\tau-\tau')\\-l^2\frac{e^{-M|\tau-\tau'|}}{2M} -l^2\frac{1}{e^{\beta M}-1}\frac{\cosh(M|\tau-\tau'|)}{M}.
    \end{split}
\end{equation}

In order to calculate $Z(A,\beta)$, the effective action is given in terms of the Matsubara modes by the expression $ S_{eff}^\beta = \frac{1}{2}\sum_j \left(\omega_j^2+M^2-\frac{l^2}{\omega_j^2+M^2}\right)x_j^*x_j$, and we can perform another Gaussian integral to arrive at,
\begin{equation}
    \log Z(\beta) = -\frac{1}{2}\sum_j\log(\omega_j^2+M^2-\frac{l^2}{\omega_j^2+M^2}).
\end{equation}

Decomposing the logarithm and using the Matsubara sum $ \sum_j\log(\omega_j^2+C^2) = 2\log\sinh(\frac{\beta C}{2})$,
\begin{equation}
    \begin{split}
        \log\left(\omega_j^2+M^2-\frac{l^2}{\omega_j^2+M^2}\right) = \log\left(\omega_j^2+M^2-l\right)\\
        +\log\left(\omega_j^2+M^2+l\right)
     -\log\left(\omega_j^2+M^2\right),
    \end{split}
\end{equation}
and so,
\begin{equation}
    Z(A,\beta) = \frac{\sinh(\frac{\beta M}{2})}{2\sinh(\frac{\beta\sqrt{M^2+l}}{2})\sinh(\frac{\beta\sqrt{M^2-l}}{2})}.
\end{equation}
The extra factor of $2$ in the denominator does not change any physical expectation value, but it allows the reduced density matrix to be properly normalized.\bigskip

Now, as shown in Section \ref{measures}, for calculating $Z_n(A, \beta)$ the local terms of the $S_{eff}$ in $\sum_{u=0}^{n-1}\int_{u\beta}^{(u+1)\beta} d\tau L^\beta_\mu$ simply add up to the same expression at inverse temperature $n\beta$, so the focus now is on the nonlocal part after taking the variables as periodic in $\beta$,
\begin{equation}
\label{intexp}
    \begin{split}
        \frac{1}{n\beta}\sum_{j,j'}x_jx_{j'}\int_0^{n\beta} d\tau\int_0^{n\beta} d\tau'\Theta_{n}(\tau,\tau')e^{i\omega_j\tau+i\omega_{j'}\tau'}\\ \times l^2\left[\frac{e^{-M|\tau-\tau'|}}{2M}+\frac{1}{e^{\beta M}-1}\frac{\cosh(M|\tau-\tau'|)}{M}\right].
    \end{split}
\end{equation}

Here it is important to make clear that the Fourier coefficients of the variables are normalized as  $x_A(\tau) = \frac{1}{\sqrt{n\beta}}\sum_je^{i\omega_j\tau}x_j$ (the same choice will be maintained in the field theory case). Furthermore, $\omega_j = \frac{2\pi j}{n\beta}$.\bigskip

For the next step, given the definition of the hyperbolic functions, the integrals over $\tau'$ and $\tau$ in equation (\ref{intexp}) only involve exponentials. So, for the factors with $e^{-M|\tau-\tau'|}$, they result in,
\begin{equation}
\label{exponential}
    \begin{split}
          \frac{1}{n\beta}\sum_{u=0}^{n-1}e^{2\pi i \frac{j+j'}{n}u}\{
        \frac{2M}{M^2+\omega^2_{j'}}\frac{e^{2\pi i \frac{j+j'}{n}}-1}{i(\omega_j+\omega_{j'})}+\\ \frac{e^{-\beta M+2\pi i \frac{j}{n}}-1}{(M+i\omega_{j'})(M-i\omega_j)}+\frac{e^{-\beta M+2\pi i \frac{j'}{n}}-e^{2\pi i \frac{j+j'}{n}}}{(M-i\omega_{j'})(M+i\omega_j)}\}. 
    \end{split}
\end{equation}
The corresponding expression obtained from the $e^{M|\tau-\tau'|}$ term appearing in the hyperbolic cosine is obviously derived from the equation above by changing the sign of $M$.

Using the identities $\sum_{u=0}^{n-1}e^{2\pi i \frac{j+j'}{n}u} = n\sum_\nu\delta_{j+j'}^{n\nu}$ and $\delta_{j+j'}^{n\nu}\frac{e^{2\pi i \frac{j+j'}{n}}-1}{i\frac{j+j'}{n}}= \delta_{j+j'}^0$, Eq.  (\ref{exponential}) is further simplified to,
\begin{equation}
\label{exponential2}
    \begin{split}
         \delta_{j+j'}^0\frac{2M}{M^2+\omega^2_{j'}}+\frac{1}{n\beta}n\sum_\nu\delta_{j+j'}^{n\nu}\times\\ \left[\frac{e^{-\beta M+2\pi i \frac{j}{n}}-1}{(M+i\omega_{j'})(M-i\omega_j)}+\frac{e^{-\beta M+2\pi i \frac{j'}{n}}-1}{(M-i\omega_{j'})(M+i\omega_j)}\right].
    \end{split}
\end{equation}

Collecting the other factors from Eq. (\ref{intexp}), there will be two main components in the new ``action" which serves to define the modified partition function $Z_n(A,\beta)$: those derived from $e^{-M|\tau-\tau'|}$ and those from $e^{M|\tau-\tau'|}$.
In the first case, we simply get Eq. (\ref{exponential2}) multiplied by $\frac{l^2}{2M}(1+\frac{1}{e^{\beta M}-1})x_jx_{j'}$ and with a sum over Matsubara frequencies $j$ and $j'$. Our interest is in the zero-temperature limit, so in this component the term $\frac{1}{e^{\beta M}-1}$
can be safely ignored as it is exponentially suppressed when $\beta\to\infty$, and so all its contributions vanish; the same can be said about the other $e^{-\beta M}$ terms inside the sum.
This means that the contribution of this component is,
\begin{equation}
\label{component1}
   \begin{split}
       \frac{1}{2}\sum_j\frac{l^2}{M^2+\omega_j^2}x^*_jx_j-n\frac{l^2}{2M}\frac{1}{n\beta}\times\\\sum_{j,\nu}\Re\frac{1}{(M+i\omega_{n\nu}-i\omega_j)(M-i\omega_j)}x_jx_{n\nu-j}.
   \end{split}
\end{equation}
Note that the first term is exactly the same as in the calculation of $Z(A,\beta)$ for inverse temperature $n\beta$. Later we will show it is responsible for canceling the denominator in the equation for $\Tr\rho^n_A$.\bigskip 

Now, moving to the second component, derived from $e^{M|\tau-\tau'|}$, the change in sign means that its corresponding version of (\ref{exponential2}) will have exponentially increasing terms $e^{\beta M}$. This combined with the overall $(e^{\beta M}-1)^{-1}$ multiplying it means that the only terms which may be relevant as $\beta\to\infty$ are given by,
\begin{equation}
   \begin{split}
        n\frac{l^2}{2M}\frac{1}{n\beta}\sum_{j,j'}\sum_\nu\delta_{j+j'}^{n\nu}x_jx_{j'}\times\\ \frac{e^{2\pi i \frac{j}{n}}}{(M+i\omega_{j'})(M-i\omega_j)}+\frac{e^{2\pi i \frac{j'}{n}}}{(M-i\omega_{j'})(M+i\omega_j)}.
   \end{split}
\end{equation}
We are doing only a lowest-order perturbative calculation. Thus, in order to see how this component contributes to $Z_n(A,\beta)$, we can expand the exponential containing it and calculate the simple path integral,
\begin{equation}
    \begin{split}
        -n\frac{l^2}{2M}\frac{1}{n\beta}\sum_{j,j',\nu}\delta_{j+j'}^{n\nu}\int\mathcal{D}x_je^{-S^{n\beta}_0}x_jx_{j'}
       \times\\\frac{e^{2\pi i \frac{j}{n}}}{(M+i\omega_{j'})(M-i\omega_j)}+\frac{e^{2\pi i \frac{j'}{n}}}{(M-i\omega_{j'})(M+i\omega_j)},
    \end{split}
\end{equation}
with the focus on the lowest order, allowing us to use the free action in the exponential, since all corrections are of higher power in $l$. 
For the discussion regarding this particular contribution we only need the sums over Matsubara frequencies and the fact that the Gaussian integral gives $\langle x_jx_{j'}\rangle =\frac{\delta_{j+j'}^0}{\omega_j^2+M^2}$. Thus, ignoring all multiplicative factors, we have,
\begin{equation}
\label{second}
    \frac{1}{n\beta}\sum_j \frac{e^{2\pi i \frac{j}{n}}}{(\omega^2_j+M^2)^2} = -\frac{d}{dM^2}\frac{1}{n\beta}\sum_j \frac{e^{2\pi i \frac{j}{n}}}{\omega^2_j+M^2}.
\end{equation}
Note that the Matsubara sum on the right-hand side is the same as in Eq. (\ref{sum}), \textit{but evaluated at time difference} $|\tau-\tau'|=\beta$ and \textit{done over frequencies associated with periodicity} $n\beta$, meaning we are left with the expression,
\begin{equation}
    -\frac{d}{dM^2}\left(\frac{e^{-M\beta}}{2M}+\frac{1}{e^{n\beta M}-1}\frac{\cosh(M\beta)}{M}\right).
\end{equation}
Therefore, in the $\beta\to\infty$ limit this entire contribution goes to zero and is irrelevant for the entanglement at this order. Note, however, that this limit only vanishes because the sum was evaluated at time $\beta$, while the frequencies were those at inverse temperature $n\beta$. Thus, the ``replica" aspect of the method, with this discrepancy in the periodicity of the traced out and remaining degrees of freedom, was essential.
Importantly, note that if there was some condition on Eq. (\ref{second}) forcing the imaginary exponential to be unity, it would be a common Matsubara sum \textit{whose zero temperature limit does not vanish} and so would contribute to $\Tr\rho_A^n$ as a positive term. We will see in the field theory cases that some components of this form (arising from the $\cosh(M|\tau-\tau'|)/(e^{\beta M}-1)$ part of the nonlocal kernel) will be such that this scenario is realized, being crucial to obtaining the correct results.\bigskip

With this, we can finally return to the contribution from Eq. (\ref{component1}). As before, we perform a perturbative expansion of the exponential and take the lowest order term. Knowing that all other contributions vanish, the modified partition function of the replica trick is, 
\begin{eqnarray}
         &&Z_n(A,\beta) = \int\mathcal{D}x_je^{-S^{n\beta}_{\rm eff}}
        \left\{1-n\frac{l^2}{2M}\frac{1}{n\beta}\right.
        \nonumber\\
        &\times&\left. \sum_{j,\nu}\Re\frac{1}{(M+i\omega_{n\nu}-i\omega_j)(M-i\omega_j)}x_jx_{n\nu-j}\right\}.
        \nonumber\\
\end{eqnarray}

As mentioned before, the effective action at inverse temperature $n\beta$ is automatically reproduced, so the order $\mathcal{O}(l^0)$ part of $Z_n(A,\beta)$ is equal to $Z(A,n\beta)$. Thus, using $\lim_{\beta\to\infty}\frac{Z(A,n\beta)}{\left[Z(A,\beta)\right]^n} = 1$, we arrive at the trace,
\begin{equation}
\begin{split}
        \Tr\rho_A^n = 1-n\frac{l^2}{2M}\lim_{\beta\to\infty}\frac{1}{n\beta}\sum_{j,\nu}\langle x_jx_{n\nu-j}\rangle\times\\\Re\frac{1}{(M+i\omega_{n\nu}-i\omega_j)(M-i\omega_j)}.
   \end{split}
\end{equation}
By Wick's theorem with average taken with respect to the effective action, $\langle x_jx_{n\nu-j}\rangle =\delta_\nu^0 (\omega^2+M^2 - \frac{l^2}{\omega^2+M^2})^{-1}$, so the sum over $\nu$ can be performed easily and the term inside the final Matsubara sum can be written as,
\begin{equation}
  \begin{split}
      \frac{4M^2}{(M^2+\omega_j^2)(M^2+\omega_j^2+l)(M^2+\omega_j^2-l)}\\-\frac{2}{(M^2+\omega_j^2+l)(M^2+\omega_j^2-l)}.
  \end{split}
\end{equation}
By using partial fraction identities, expanding the denominators in $l$ at lowest order, performing the usual Matsubara sums and taking the zero-temperature limit, we obtain,
\begin{equation}
    \Tr\rho_A^n = 1-n\frac{l^2}{16M^4}.
\end{equation}

\section{Momentum-Space entropy in $\phi^3$ theory}
\label{a2}
As explained in Section \ref{phi3} in terms of Feynman diagrams, the Wilsonian integration of fast modes in the $\phi^3$ theory leads to nonlocal terms at order $\mathcal{O}(\lambda^2)$ given by,
\begin{equation}
 \frac{1}{\omega_{j'}^2+\Vec{q}^2+m^2}\frac{1}{(\omega_j+\omega_{j'})^2+(\Vec{k}-\Vec{q})^2+m^2}|\phi_{j,\Vec{k}}|^2,
\end{equation}
with $|\Vec{k}|<\mu$, a Matsubara sum over $j'$ and an integral over $\Vec{q}$ such that $|\Vec{q}|,|\Vec{k}-\Vec{q}|>\mu$. Furthermore, the other nonlocal term is,
\begin{equation}
 \frac{(2\pi)^{d-1}\beta\delta(\sum_i\Vec{k}_i)\delta(\sum_ij_i)\phi_{j_1,\Vec{k}_1}\phi_{j_2,\Vec{k}_2}\phi_{j_3,\Vec{k}_3}\phi_{j_4,\Vec{k}_4}}{(\omega_{j_1}+\omega_{j_2})^2+(\Vec{k}_1+\Vec{k}_2)^2+m^2},
\end{equation}
such that $|\Vec{k}_i|<\mu$ while $|\Vec{k}_1+\Vec{k}_2|>\mu$.
Note that in both cases, as long as the integrals over momenta are left for the end, we may use the same calculations as in the previous example of coupled harmonic oscillators. 
Going from Matsubara modes of the fields to Euclidean time, we see that the terms above are indeed nonlocal. In order to show how this, we must first introduce a well-known Matsubara sum we will use in the remaining Appendices (see Ref. \cite{lebellac} for a derivation),
\begin{equation}
\label{doublepropagator}
\begin{split}
        \frac{1}{\beta}\sum_{j'}\frac{1}{(\omega_j+\omega_{j'})^2+E_1^2}\frac{1}{\omega_{j'}^2+E_2^2}=&\\(1+n(E_1)+n(E_2))\frac{E_1+E_2}{2E_1E_2}\frac{1}{\omega_j^2+(E_1+E_2)^2}\\+(n(E_1)-n(E_2))\frac{E_2-E_1}{2E_1E_2}\frac{1}{\omega_j^2+(E_2-E_1)^2},
    \end{split}
\end{equation}
where $n(E)$ denotes the Bose-Einstein distribution.

In the $\beta\to\infty$ limit, the Bose-Einstein terms are suppressed exponentially even before the Fourier transform is performed. Thus, such terms do not contribute to the entropy and can be safely ignored. Having this point in mind and using Eq. (\ref{sum}), the relevant nonlocal kernel of the two-field term obtained by Fourier transforming only the time components of the fields is, 
\begin{equation}
\label{phi3twofield}
 \frac{e^{-M|\tau-\tau'|}+(e^{\beta M}-1)^{-1}e^{M|\tau-\tau'|}}{4\sqrt{\Vec{q}^2+m^2}\sqrt{(\Vec{k}-\Vec{q})^2+m^2}},
\end{equation}
where the unimportant terms were excluded and the new decay rate of the exponentials is $M = \sqrt{\Vec{q}^2+m^2}+\sqrt{(\Vec{k}-\Vec{q})^2+m^2}$.

In the four-field term, we can prove the nonlocality by first writing it in a generic form,
\begin{equation}
 \frac{1}{\beta^2}\sum_{j_1,j_2,j_3}\phi_{1,j_1}\phi_{2,j_2}f(\omega_{j_1}+\omega_{j_2})\phi_{3,j_3}\phi_{4,-j_1-j_2-j_3},
\end{equation}
with all multiplicative constants suppressed and the dependence on spatial momenta is represented by the numerical indices in the fields. The advantage of writing the term so generically is that the final result will automatically be valid for the four- and six-field terms in the $\phi^4$ case with minor modifications. 

Writing the fields in Euclidean time, the expression becomes,
\begin{equation}
    \label{nonlocalderivation}
    \begin{split}
    \frac{1}{\beta^2}\sum_{j_1,j_2,j_3}\int_0^\beta d\tau_1d\tau_2d\tau_3d\tau_4\phi_1(\tau_1)\phi_2(\tau_2)\phi_3(\tau_3)\phi_4(\tau_4)\\\times
e^{i\omega_{j_1}(\tau_1-\tau_4)+i\omega_{j_2}(\tau_2-\tau_4)+i\omega_{j_3}(\tau_3-\tau_4)} f(\omega_{j_1}+\omega_{j_2}) \\= \frac{1}{\beta}\sum_{j_1,j_2}\int_0^\beta d\tau_1d\tau_2d\tau_3\phi_1(\tau_1)\phi_2(\tau_2)\phi_3(\tau_3)\phi_4(\tau_3)\\\times
e^{i(\omega_{j_1}+\omega_{j_2})(\tau_1-\tau_3)+i\omega_{j_2}(\tau_2-\tau_1)} f(\omega_{j_1}+\omega_{j_2}) \\
= \int_0^\beta d\tau\int_0^\beta d\tau'\phi_1(\tau)\phi_2(\tau)\Tilde{f}(\tau-\tau')\phi_3(\tau')\phi_4(\tau'),
\end{split}
\end{equation}
where $\Tilde{f}(\tau)$ is the Fourier transform of $f(\omega_j)$. 

For our specific case in this Appendix, this means the nonlocal kernel of the four-field term is given by eq. (\ref{sum}) with $M=\sqrt{(\Vec{k}_1+\Vec{k}_2)^2+m^2}$. Furthermore, when we apply the replica trick, the structure of the equation above is such that calculating the sum of double integrals will proceed as in the previous section, the only difference being the replacement of the single Matsubara frequency $\omega_j$ by the sum $\omega_{j_1}+\omega_{j_2}$ and of $\omega_{j'}$ by $\omega_{j_3}+\omega_{j_4}$ in the imaginary exponents.\bigskip

The modified partition function $Z_n(\mu,\beta)$ can now be calculated up to order $\mathcal{O}(\lambda^2)$ and since both nonlocal terms are of the same form as in the case of coupled oscillators, the sum of double integrals over $\tau$ and $\tau'$ can be calculated by adapting eq. (\ref{exponential}), taking care to use the new expressions for $M$ and the correct multiplicative factors. 

In more detail, the expression in terms of the Matsubara frequencies for the two and four-field terms become, respectively,
\begin{equation}
    \begin{split}
      \sum_j\frac{2M}{\omega_j^2+M^2}\phi^*_{j,\Vec{k}}\phi_{j,\Vec{k}}-n\frac{1}{n\beta}\times\\\sum_{j,\nu}2\Re\frac{1}{(M+i\omega_{n\nu}-i\omega_j)(M-i\omega_j)}\phi_{j,\Vec{k}}\phi_{n\nu-j,-\Vec{k}},
    \end{split}
\end{equation}
which must be multiplied by $\frac{1}{4\sqrt{\Vec{q}^2+m^2}\sqrt{(\Vec{k}-\Vec{q})^2+m^2}}$ before including the remaining momentum integrals and numerical factors and with $M=\sqrt{\Vec{q}^2+m^2}+\sqrt{(\Vec{k}-\Vec{q})^2+m^2}$ as pointed out earlier, and,
\begin{equation}
    \begin{split}
        \frac{1}{\beta}\sum_{j_1,j_2,j_3,j_4}\frac{2M\delta(\sum_ij_i)}{(\omega_{j_1}+\omega_{j_2})^2+M^2}\phi_{j_1,\Vec{k}_1}\phi_{j_2,\Vec{k}_2}\phi_{j_3,\Vec{k}_3}\phi_{j_4,\Vec{k}_4}\\
       -n\frac{1}{(n\beta)^2}\sum_{j_1,j_2,j_3,j_4,\nu}\delta^{n\nu}_{j_1+j_2+j_3+j_4}\phi_{j_1,\Vec{k}_1}\phi_{j_2,\Vec{k}_2}\phi_{j_3,\Vec{k}_3}\phi_{j_4,\Vec{k}_4}\\\times2\Re\frac{1}{M+i(\omega_{j_3}+\omega_{j_4})}\frac{1}{M-i(\omega_{j_1}+\omega_{j_2})},
    \end{split}
\end{equation}
with $M=\sqrt{(\Vec{k}_1+\Vec{k}_2)^2+m^2}$ and multiplied by the remaining factors, which include $(2\pi)^{d-1}\delta(\sum_{i=1}^4\Vec{k}_i)$ (and, of course, integrating over the proper momentum regions indicated each diagram).\bigskip

Once again there are terms identical to those in $Z(\mu,n\beta)$, meaning they are canceled in the entropy when taking the zero temperature limit (since $Z(\mu,n\beta)$ is equivalent to $Z(\mu,\beta)^n$ as $\beta\to\infty$). With this, the Rényi entropies are simply given by the remaining terms divided by a $Z_0(\mu,\beta)^n$ factor which, again using the equality of partition function limits, can be replaced by $Z_0(\mu,n\beta)$ and leads to expectation values of products of fields.

Thus, the lowest-order Rényi entropy will depend on the following Matsubara sums:
\begin{equation}
\label{avg2}
 \lim_{\beta\to\infty}\frac{1}{n\beta}\sum_{j,\nu}2\Re\frac{\langle\phi_{j,\Vec{k}}\phi_{n\nu-j,-\Vec{k}}\rangle}{(M+i\omega_{n\nu}-i\omega_j)(M-i\omega_j)},
\end{equation}
\begin{equation}
 \label{avg4}
 \begin{split}
       \lim_{\beta\to\infty}\frac{2}{(n\beta)^2}\sum_{j_1,j_2,j_3,j_4,\nu}\langle\phi_{j_1,\Vec{k}_1}\phi_{j_2,\Vec{k}_2}\phi_{j_3,\Vec{k}_3}\phi_{j_4,\Vec{k}_4}\rangle\times\\\delta^{n\nu}_{j_1+j_2+j_3+j_4}\Re\frac{1}{M+i(\omega_{j_3}+\omega_{j_4})}\frac{1}{M-i(\omega_{j_1}+\omega_{j_2})}.
 \end{split}
\end{equation}
The field averages are given by Wick's theorem, so we have the possible contractions of field products and for each contraction $\langle\phi_{j,\Vec{k}}\phi_{j',\Vec{p}}\rangle = \delta_j^{j'}\frac{(2\pi)^{d-1}\delta(\Vec{k}-\Vec{p})}{\omega_j^2+\Vec{k}^2+m^2}$. Just as in ordinary free energy calculations, the contractions lead to the presence of a delta function $\delta(\Vec{k}-\Vec{k})$ which is not well defined. We then consider the theory in a volume $V$ and have: $(2\pi)^{d-1}\delta(\Vec{k}-\Vec{k}) = \int d^{d-1}x e^{i\Vec{x} (\Vec{k}-\Vec{k})} = V$, so the entropy will be an extensive quantity as discussed previously. 

For the two-field term there is only one possible contraction and for this contraction we can follow verbatim the steps made in the previous Appendix to show that the contribution from the $\frac{e^{M|\tau-\tau'|}}{e^{\beta M}-1}$ term of the nonlocal kernel vanishes exponentially in the zero temperature limit just as in the coupled harmonic oscillator case. Thus, eq. (\ref{avg2}) is the only relevant part of the two-field term and, after a number of algebraic manipulations and Matsubara sums, we find that its zero temperature limit is,
\begin{equation}
\label{expression}
    \frac{1}{\sqrt{\Vec{k}^2+m^2}\left(M+\sqrt{\Vec{k}^2+m^2}\right)^2},
\end{equation}
such that $M = \sqrt{\Vec{q}^2+m^2}+\sqrt{(\Vec{k}-\Vec{q})^2+m^2}$.

Likewise, for the four-field term, there are three possible ways of contracting the product, two of which are equal. As discussed in Section \ref{phi3}, the contraction corresponding to diagram,
%
\begin{figure}[H]
	\centering
	\includegraphics[width=4cm]{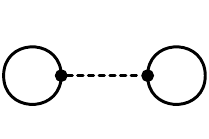}
\end{figure}
\noindent
is identically zero due to conflicting momentum restrictions.

For the remaining possibilities, their structure is such that, similarly to the two-field term, contributions from $\frac{e^{M|\tau-\tau'|}}{e^{\beta M}-1}$ vanish (as can be seen by carrying them throughout the calculation) and so, after lengthy but simple calculations, we find that the contribution of the four-field term is,
\begin{equation}
\label{expression2}
    \frac{1}{\sqrt{\Vec{k}_1^2+m^2}\sqrt{\Vec{k}_2^2+m^2}\left(M+\sqrt{\Vec{k}_1^2+m^2}+\sqrt{\Vec{k}_2^2+m^2}\right)^2}.
\end{equation}
with $M=\sqrt{(\Vec{k}_1+\Vec{k}_2)^2+m^2}$.\bigskip

Finally, the lowest-order entropy is simply the sum of both contributions with momentum integrals and multiplicative factors restored (note that they arise from diagrams of similar structure and have the same integrands). As mentioned in the main text, in order to compare our result with that of Ref. \cite{Balasubramanian:2011wt}, we count the possible permutations of high and low momenta (and multiply each contribution by the appropriate factor) and restrict the integration regions accordingly. Therefore, our final result for the Rényi entropy at lowest order of the $\phi^3$ theory is given by,
\begin{equation}
    \begin{split}
        \frac{ H_n(\mu)}{V} = \frac{n}{n-1} \frac{\lambda^2}{8}\int^*\frac{d^{d-1}k}{(2\pi)^{d-1}}\frac{d^{d-1}q}{(2\pi)^{d-1}} \mathcal{I}(\Vec{k}, \Vec{q}),
    \end{split}
\end{equation}
with,
\begin{equation}
  \begin{split}
      \mathcal{I}(\Vec{k}, \Vec{q})\equiv  \frac{1}{\sqrt{\Vec{k}^2+m^2}}\frac{1}{\sqrt{\Vec{q}^2+m^2}}
    \frac{1}{\sqrt{(\Vec{k}-\Vec{q})^2+m^2}}\\
    \times
     \frac{1}{\left(\sqrt{\Vec{k}^2+m^2}+\sqrt{\Vec{q}^2+m^2}+\sqrt{(\Vec{k}-\Vec{q})^2+m^2}\right)^2},
  \end{split}
\end{equation}
and the integration region being (as a consequence of the momentum restrictions of the diagrams which contribute and the elimination of permutations we made) such that no set of momenta $\Vec{k}, \Vec{q}, \Vec{k}-\Vec{q}$ is repeated and at least one of the three is above scale $\mu$ and at least one is below it. 

\section{Momentum-Space entropy in $\phi^4$ theory}
\label{a3}

For the $\phi^4$ calculation we can draw a lot from the derivations made in the previous two Appendices. To do so, we first write the finite temperature expressions associated with the relevant diagrams discussed in Section (\ref{phi4}),
\begin{equation}
 \begin{aligned}
 \frac{1}{\beta^3}\sum_{j,j_1,j_2}\frac{1}{\omega_{j_1}^2+\Vec{q}^2+m^2}\frac{1}{\omega_{j_2}^2+\Vec{p}^2+m^2}\times\\
 \frac{1}{(\omega_j+\omega_{j_1}+\omega_{j_2})^2+(\Vec{k}+\Vec{q}+\Vec{p})^2+m^2}|\phi_{j,\Vec{k}}|^2,
 \end{aligned}
\end{equation}

\begin{equation}
    \begin{split}
      \frac{1}{\beta^3}\sum_j \frac{1}{(\omega_{j_1}+\omega_{j_2}+\omega_j)^2+(\Vec{k}_1+\Vec{k}_2-\Vec{q})^2+m^2}\\
       \frac{1}{\omega_j^2+\Vec{q}^2+m^2}2\pi\delta(\sum_{i=1}^4(\Vec{k}_i))\phi_{j_1,\Vec{k}_1}\phi_{j_2,\Vec{k}_2}\phi_{j_3,\Vec{k}_3}\phi_{j_4,\Vec{k}_4},
    \end{split}
\end{equation}
\begin{equation}
\begin{aligned}
 \frac{1}{\beta^3}\frac{(2\pi)^{d-1}\delta(\sum_{i=1}^6\Vec{k}_i)}{(\omega_{j_1}+\omega_{j_2}+\omega_{j_3})^2+(\Vec{k}_1+\Vec{k}_2+\Vec{k}_3)^2+m^2}\\
\times\delta(\sum_{i=1}^6j_i)\phi_{j_1,\Vec{k}_1}\phi_{j_2,\Vec{k}_2}\phi_{j_3,\Vec{k}_3}\phi_{j_4,\Vec{k}_4}\phi_{j_5,\Vec{k}_5}\phi_{j_6,\Vec{k}_6},
\end{aligned}
\end{equation}
with some factors and integrals suppressed for convenience.

To find how these terms are nonlocal in Euclidean time, we use equations (\ref{sum}) and (\ref{doublepropagator}) and we also need to employ (and adapt) the derivation (\ref{nonlocalderivation}) to see that given \textit{the specific structure of the Feynman diagrams} generating such terms, the four-field term will be of the form, 
\begin{equation}
  \phi_{\Vec{k}_1}(\tau)\phi_{\Vec{k}_2}(\tau)f(\tau-\tau')\phi_{\Vec{k}_3}(\tau')\phi_{\Vec{k}_4}(\tau')  
\end{equation} 
and the six-field one will be, 
\begin{equation}
\phi_{\Vec{k}_1}(\tau)\phi_{\Vec{k}_2}(\tau)\phi_{\Vec{k}_3}(\tau)g(\tau-\tau')\phi_{\Vec{k}_4}(\tau')\phi_{\Vec{k}_5}(\tau')\phi_{\Vec{k}_6}(\tau').    
\end{equation}
In more detail, we have seen that after Matsubara sums and the Fourier transform, the nonlocal term is (before momentum integrals) an exponential function and will be of the forms,
\begin{equation}
\frac{e^{-(\sqrt{\Vec{q}^2+m^2}+\sqrt{\Vec{p}^2+m^2}+\sqrt{(\Vec{k}-\Vec{q}-\Vec{p})^2+m^2})|\tau-\tau'|}}{8\sqrt{\Vec{q}^2+m^2}\sqrt{\Vec{p}^2+m^2}\sqrt{(\Vec{k}-\Vec{q}-\Vec{p})^2+m^2}}\phi_{\Vec{k}}(\tau)\phi_{\Vec{k}}^*(\tau'),
\end{equation}
\begin{equation}
   \begin{split}
        \frac{e^{-(\sqrt{\Vec{q}^2+m^2}+\sqrt{(\Vec{k}_1+\Vec{k}_2-\Vec{q})^2+m^2})|\tau-\tau'|}}{4\sqrt{\Vec{q}^2+m^2}\sqrt{(\Vec{k}_1+\Vec{k}_2-\Vec{q})^2+m^2}}\\\times\phi_{\Vec{k}_1}(\tau)\phi_{\Vec{k}_2}(\tau)\phi_{\Vec{k}_3}(\tau')\phi_{\Vec{k}_4}(\tau'),
   \end{split}
\end{equation}
\begin{equation}
    \begin{aligned}
     \frac{e^{-\sqrt{(\Vec{k}_1+\Vec{k}_2+\Vec{k}_3)^2+m^2}|\tau-\tau'|}}{2\sqrt{(\Vec{k}_1+\Vec{k}_2+\Vec{k}_3)^2+m^2}}\\
     \times\phi_{\Vec{k}_1}(\tau)\phi_{\Vec{k}_2}(\tau)\phi_{\Vec{k}_3}(\tau)\phi_{\Vec{k}_4}(\tau')\phi_{\Vec{k}_5}(\tau')\phi_{\Vec{k}_6}(\tau'). 
    \end{aligned}
\end{equation}
Thus, to apply the replica trick we have the general form of the kernel $e^{-M|\tau-\tau'|}$ with decay rates $M= \sqrt{\Vec{q}^2+m^2}+\sqrt{\Vec{p}^2+m^2}+\sqrt{(\Vec{k}-\Vec{q}-\Vec{p})^2+m^2}$, $\Tilde{M} = \sqrt{\Vec{q}^2+m^2}+\sqrt{(\Vec{k}_1+\Vec{k}_2-\Vec{q})^2+m^2}$ and $\hat{M}=\sqrt{(\Vec{k}_1+\Vec{k}_2+\Vec{k}_3)^2+m^2}$ for terms with two, four and six terms, respectively. It is important to remember that besides the expressions written above, there are also the ones associated with $\frac{\cosh{M|\tau-\tau'|}}{M(e^{\beta M}-1)}$ which also appear from the Fourier transform.\bigskip

To obtain the contributions of each nonlocal expression to the entropy, many of the steps of the $\phi^3$ calculation can be followed verbatim. Denoting by $H_n^{(2)}(\mu)$ the contribution from the two-field term, we just need to use the new expression for $M$ in equation (\ref{expression}) and arrive at,
\begin{equation}
    \frac{H_n^{(2)}(\mu)}{V} = \frac{n}{n-1} \frac{\lambda^2}{96}\int^*\prod_{i=1}^3\frac{d^{d-1}k_i}{(2\pi)^{d-1}}\mathcal{I}(\Vec{k}_1,\Vec{k}_2,\Vec{k}_3),
\end{equation}
\begin{equation}
\begin{split}
\mathcal{I}(\Vec{k}_1,\Vec{k}_2,\Vec{k}_3) \equiv \frac{1}{\sqrt{(\sum_{i=1}^3\Vec{k}_i)^2+m^2}}\prod_{i=1}^3\frac{1}{\sqrt{\Vec{k}_i^2+m^2}}\\
 \times \frac{1}{\left(\sum_{i=1}^3\sqrt{\Vec{k}_i^2+m^2}+\sqrt{(\sum_{i=1}^3\Vec{k}_i)^2+m^2}\right)^2},
    \end{split}
\end{equation}
with $|\Vec{k}_1|<\mu$ and $|\Vec{k}_2|,|\Vec{k}_3|,|\Vec{k}_1+\Vec{k}_2+\Vec{k}_3|>\mu$.\bigskip

To deal with the four-field term it's important to calculate again $\sum_{u=0}^{n-1}\int_{u\beta}^{(u+1)\beta} d\tau\int_{u\beta}^{(u+1)\beta} d\tau'\Tilde{L}(\tau,\tau')$. As before, we do the calculation for an exponential kernel and the result is easily generalized for actual finite temperature appearing. Because of the way the imaginary times of the fields are paired, this sum becomes,
\begin{equation}
   \begin{split}
         \frac{1}{n\beta}\sum_{j_1,j_2,j_3,j_4}\frac{2\Tilde{M}\delta^0_{j_1+j_2+j_3+j_4}}{\Tilde{M}^2+(\omega_{j_3}+\omega_{j_4})^2}\phi_{j_1,\Vec{k}_1}\phi_{j_2,\Vec{k}_2}\phi_{j_3,\Vec{k}_3}\phi_{j_4,\Vec{k}_4}\\
    -\frac{2n}{(n\beta)^2}\sum_{j_1,j_2,j_3,j_4}\delta^{n\nu}_{j_1+j_2+j_3+j_4}\phi_{j_1,\Vec{k}_1}\phi_{j_2,\Vec{k}_2}\phi_{j_3,\Vec{k}_3}\phi_{j_4,\Vec{k}_4}\\\times\Re\frac{1}{\Tilde{M}+i(\omega_{j_3}+\omega_{j_4})}\frac{1}{\Tilde{M}-i(\omega_{j_1}+\omega_{j_2})}.
   \end{split}
\end{equation}
Once again, the contribution from the first term will be cancelled when calculating the entropy and we are left with the sum,
\begin{equation}
\begin{split}
    \sum_{j_1,j_2,j_3,j_4}\frac{\delta^{n\nu}_{j_1+j_2+j_3+j_4}}{(n\beta)^2}\langle\phi_{j_1,\Vec{k}_1}\phi_{j_2,\Vec{k}_2}\phi_{j_3,\Vec{k}_3}\phi_{j_4,\Vec{k}_4}\rangle \\\times\Re\frac{1}{\Tilde{M}+i(\omega_{j_3}+\omega_{j_4})}\frac{1}{\Tilde{M}-i(\omega_{j_1}+\omega_{j_2})}.
\end{split}   
\end{equation}

The average $\langle\phi_{j_1,\Vec{k}_1}\phi_{j_2,\Vec{k}_2}\phi_{j_3,\Vec{k}_3}\phi_{j_4,\Vec{k}_4}\rangle$ is calculated via Wick's theorem and each contraction is given $\langle\phi_{j_1,\Vec{k}_1}\phi_{j_2,\Vec{k}_2}\rangle = \delta_{j_1+j_2}^0\frac{(2\pi)^{d-1}\delta(\Vec{k}_1-\Vec{k}_2)}{\omega_j^2+\Vec{k}_1^2+m^2}$. 

This is the point at which the possible contractions give rise to the associated ``basketball" and ``cactus" diagrams. The calculation for the basketball, whose contribution we denote $H_n^{(4)}(\mu)$ is a matter of long but straightforward algebraic manipulations, similar to those of the nonlocal four-field term in the $\phi^3$ theory, and it culminates (remembering to use the expression for $\Tilde{M}$) in,
\begin{equation}
  \frac{H_n^{(4)}(\mu)}{V}= \frac{n}{n-1} \frac{\lambda^2}{64}\int^*\prod_{i=1}^3 \frac{d^{d-1}k_i}{(2\pi)^{d-1}}\mathcal{I}(\Vec{k}_1,\Vec{k}_2,\Vec{k}_3),
\end{equation}
with $|\Vec{k}_1|,|\Vec{k}_2|\leq\mu$, $\mu\leq|\Vec{k}_3|,|\Vec{k}_1+\Vec{k}_2+\Vec{k}_3|$.\bigskip

Now, note that the analogous of eq.(\ref{second}) appears in the case of the four-field term but with the replacement $j=j_1+j_2$, this means that for the cactus diagram the field contractions make $j_1+j_2=0$
and so this is the specific case in which the contributions coming from the hyperbolic cosine part of the kernel \textit{don't vanish by themselves}. Furthermore, it's easy to see from eqs. (\ref{intexp}), (\ref{exponential}) and (\ref{second}) that this term as \textit{exactly same factors and opposite sign} than the cactus contribution from the decreasing exponential. Thus, by its very structure, this type of term is automatically canceled when applying the replica trick and so only ``basketballs" contribute to the entropy.

Moving to the contractions of the $\phi^6$ term, again only the ``basketballs", whose contribution we denote $H_n^{(6)}(\mu)$ are relevant. The actual calculation follows along the same lines shown throughout the previous Sections and Appendices and it is mostly busy work involving Matsubara sums and partial fraction manipulations. At the end of all steps we arrive at,
\begin{equation}
  \frac{H_n^{(6)}(\mu)}{V}= \frac{n}{n-1} \frac{\lambda^2}{96}\int^*\prod_{i=1}^3 \frac{d^{d-1}k_i}{(2\pi)^{d-1}}\mathcal{I}(\Vec{k}_1,\Vec{k}_2,\Vec{k}_3),
\end{equation}
with $|\Vec{k}_1|,|\Vec{k}_2|, |\Vec{k}_3|\leq\mu$, $\mu\leq|\Vec{k}_1+\Vec{k}_2+\Vec{k}_3|$.\bigskip

Finally, the complete result is $H_n(\mu)= H_n^{(2)}(\mu)+H_n^{(4)}(\mu)+H_n^{(6)}(\mu)$ and before performing this sum we restrict the integration regions (which we have been carrying implicitly throughout the steps) of each term and multiply them by the number of permutations of lines of same type ($3!$, $2\times2$ and $3!$, respectively), this makes the numerical factors are all equal and the overall sum becomes precisely the expression in eq. (\ref{entropyphi4}), as previously claimed.

\bibliography{citations}

\end{document}